\def\CMP{\sevenrm Commun.\ Math.\ Phys.}
\def\JMP{\sevenrm J.\ Math.\ Phys.}

\def\RMP{\sevenrm Rev.\ Math.\ Phys.}
%
%            DATUM
%
\def\today{\number\day .\space\ifcase\month\or
January\or February\or March\or April\or May\or June\or
July\or August\or September\or October\or November\or December\fi, \number \year}
%
%           Theorem, Proposition, Corollary, Lemma, Conjecture
%
\newcount \theoremnumber
\def\cleartheoremnumber{\theoremnumber = 0 \relax}

\def\Prop #1 {
             \advance \theoremnumber by 1
             \vskip .6cm 
             \goodbreak 
             \noindent
             {\bf Proposition {\the\headlinenumber}.{\the\theoremnumber}.}
             {\sl #1}  \goodbreak \vskip.8cm}

\def\Conj#1 {
             \advance \theoremnumber by 1
             \vskip .6cm  
             \goodbreak 
             \noindent
             {\bf Conjecture {\the\headlinenumber}.{\the\theoremnumber}.}
             {\sl #1}  \goodbreak \vskip.8cm} 

\def\Th#1 {
             \advance \theoremnumber by 1
             \vskip .6cm  
             \goodbreak 
             \noindent
             {\bf Theorem {\the\headlinenumber}.{\the\theoremnumber}.}
             {\sl #1}  \goodbreak \vskip.8cm}

\def\Lm#1 {
             \advance \theoremnumber by 1
             \vskip .6cm  
             \goodbreak 
             \noindent
             {\bf Lemma {\the\headlinenumber}.{\the\theoremnumber}.}
             {\sl #1}  \goodbreak \vskip.8cm}

\def\Cor#1 {
             \advance \theoremnumber by 1
             \vskip .6cm  
             \goodbreak 
             \noindent
             {\bf Corollary {\the\headlinenumber}.{\the\theoremnumber}.}
             {\sl #1}  \goodbreak \vskip.8cm} 
%
%            Formelnummerierung
%
\newcount \equationnumber

\newcount \refnumber

\def\[]    {\global 
            \advance \refnumber by 1
            [{\the\refnumber}]}

\def\# #1  {\global 
            \advance \equationnumber by 1
            $$ #1 \eqno ({\the\equationnumber}) $$ }

\def\% #1 { \global
            \advance \equationnumber by 1
            $$ \displaylines{ #1 \hfill \llap ({\the\equationnumber}) \cr}$$} 

\def\& #1 { \global
            \advance \equationnumber by 1
            $$ \eqalignno{ #1 & ({\the\equationnumber}) \cr}$$}
%
%                             Referencen
%                                
\newcount \Refnumber

\def\Ref #1 #2 #3 #4 #5 #6  {\ninerm \global
                             \advance \Refnumber by 1
                             {\ninerm #1,} 
                             {\ninesl #2,} 
                             {\ninerm #3.} 
                             {\ninebf #4,} 
                             {\ninerm #5,} 
                             {\ninerm (#6)}\nobreak} 
\def\Bookk #1 #2 #3 #4       {\ninerm \global
                             \advance \Refnumber by 1
                             {\ninerm #1,}
                             {\ninesl #2,} 
                             {\ninerm #3,} 
                             {(#4)}}
\def\Book{\cr
{\the\Refnumber} &
\Bookk}
\def\Reff{\cr
{\the\Refnumber} &
\Ref}
\def\REF #1 #2 #3 #4 #5 #6 #7   {{\sevenbf [#1]}  & \hskip -9.5cm \vtop {
                                {\sevenrm #2,} 
                                {\sevensl #3,} 
                                {\sevenrm #4} 
                                {\sevenbf #5,} 
                                {\sevenrm #6} 
                                {\sevenrm (#7)}}\cr}
\def\BOOK #1 #2 #3 #4  #5   {{\sevenbf [#1]}  & \hskip -9.5cm \vtop {
                             {\sevenrm #2,}
                             {\sevensl #3,} 
                             {\sevenrm #4,} 
                             {\sevenrm #5.}}\cr}
\def\HEP #1 #2 #3 #4     {{\sevenbf [#1]}  & \hskip -9.5cm \vtop {
                             {\sevenrm #2,}
                             {\sevensl #3,} 
                             {\sevenrm #4.}}\cr}
%
%                               Beweis
%
\def\bull{$\sqcup \kern -0.645em \sqcap$}
%
%               Definition, Remark, Example, Proof  
%
\def\Def#1{\vskip .3cm \goodbreak \noindent
                                     {\bf Definition.} #1 \goodbreak \vskip.4cm}
\def\Rem#1{\vskip .4cm \goodbreak \noindent
                                     {\it Remark.} #1 \goodbreak \vskip.5cm }

\def\Not#1{\vskip .4cm  \goodbreak \noindent
                                     {\it Notation.} #1 \goodbreak \vskip.5cm }

\def\Pr#1{\goodbreak \noindent {\it Proof.} #1 \hfill \bull  \goodbreak \vskip.5cm}

%
%                               Abstand
%
\def\*{\vskip 1.0cm}      

%
%            †berschriften
%
\newcount \ssubheadlinenumber

\def\SSHL #1 {\goodbreak
            \cleartheoremnumber
            \vskip 1cm
            \advance \ssubheadlinenumber by 1
{\rm \noindent {\the\headlinenumber}.{\the\subheadlinenumber}.{\the\ssubheadlinenumber}. #1}
            \nobreak \vskip.8cm \rm \noindent}
\newcount \subheadlinenumber
\def\clearsubheadlinenumber{\subheadlinenumber = 0 \relax}
\def\SHL #1 {\goodbreak
            \cleartheoremnumber
            \vskip 1cm
            \advance \subheadlinenumber by 1
            {\rm \noindent {\the\headlinenumber}.{\the\subheadlinenumber}. #1}
            \nobreak \vskip.8cm \rm \noindent}
\newcount \headlinenumber

\newcount \headlinesubnumber
\def\clearheadlinesubnumber{\headlinesubnumber = 0 \relax}
\def\Hl #1 {\goodbreak
            \cleartheoremnumber
            \clearheadlinesubnumber
            \clearsubheadlinenumber
            \advance \headlinenumber by 1
            {\bf \noindent {\the\headlinenumber}. #1}
            \nobreak \vskip.4cm \rm \noindent}

\font\twentyrm=cmr17
\font\fourteenrm=cmr10 at 14pt
\font\sevensl=cmsl10 at 7pt
\font\sevenit=cmti7 

\font\css=cmss10
\font\Rosch=cmr10 at 9.85pt
\font\Cosch=cmss12 at 9.5pt
\font\rosch=cmr10 at 7.00pt
\font\cosch=cmss12 at 7.00pt
\font\nosch=cmr10 at 7.00pt
%
%
%
%
%
%
%
%                   Symbols
%
%
%
%                              
%
%
%
%               reelle, nat\"urliche, ganze, komplexe Zahlen
%
%    Beispiel:  ${\bf E}_\alpha {s \atop {\raise 4pt \hbox{$\to$}}} 1$
%
\def\Z                 {\hbox{{\css Z}  \kern -1.1em {\css Z} \kern -.2em }}
\def\R                 {\hbox{\raise .03ex \hbox{\Rosch I} \kern -.55em {\rm R}}}
\def\N                 {\hbox{\rm I \kern -.55em N}}
\def\C                 {\hbox{\kern .20em \raise .03ex \hbox{\Cosch I} \kern -.80em {\rm C}}}

\def\r                 {\hbox{\raise .03ex \hbox{\rosch I} \kern -.45em \hbox{\rosch R}}}
\def\n                 {\hbox{\hbox{\rosch I} \kern -.45em \hbox{\nosch N}}}
\def\c                 {\hbox{\raise .03ex \hbox{\cosch I} \kern -.70em \hbox{\rosch C}}}

\def\z                 {\hbox{\kern 0.2em {\cal z}  \kern -0.6em {\cal z} \kern -0.3em  }}
\def\1                 {\hbox{\rm \thinspace \thinspace \thinspace \thinspace
                                  \kern -.50em  l \kern -.85em 1}}
\def\unit                 {\hbox{\sevenrm \thinspace \thinspace \thinspace \thinspace
                                  \kern -.50em  l \kern -.85em 1}}
%
%
%
%
%
%                   
%                              Sonderzeichen alg. QFT
%
%
%
%                              
%
%
\def\Tr                {{\rm Tr}}
\def\A                 {{\cal A}}

\def\B                 {{\cal B}} 

\def\H                 {{\cal H}} 
 
\def\O                 {{\cal O}}

\def\bra{\langle}
\def\ket{\rangle}

%
%                                
%
%
%                             Referencen
%                                
%
%

%\Z  
%\R  
%\N  
%\C  
%\Ro 
%\No 
%\Zo 
%\Co 
%\r  
%\n  
%\c  
%\ro 
%\no 
%\co 
%\z  
%\1  
%\unit  

\def\versuch #1 #2 {
\vskip -.1 cm
\global \advance \equationnumber by 1
            $$\displaylines{ \rlap{ #1 } \hfill #2  \hfill \llap{({\the\equationnumber})} } $$ 
\vskip  .1cm
\noindent}

\def\CMP{\sevenrm Commun.\ Math.\ Phys.}
\def\JMP{\sevenrm J.\ Math.\ Phys.}

\def\RMP{\sevenrm Rev.\ Math.\ Phys.}
%
%            DATUM
%
\def\today{\number\day .\space\ifcase\month\or
January\or February\or March\or April\or May\or June\or
July\or August\or September\or October\or November\or December\fi, \number \year}
%
%           Theorem, Proposition, Corollary, Lemma, Conjecture
%
\newcount \theoremnumber
\def\cleartheoremnumber{\theoremnumber = 0 \relax}

\def\Prop #1 {
             \advance \theoremnumber by 1
             \vskip .6cm 
             \goodbreak 
             \noindent
             {\bf Proposition {\the\headlinenumber}.{\the\theoremnumber}.}
             {\sl #1}  \goodbreak \vskip.8cm}

\def\Conj#1 {
             \advance \theoremnumber by 1
             \vskip .6cm  
             \goodbreak 
             \noindent
             {\bf Conjecture {\the\headlinenumber}.{\the\theoremnumber}.}
             {\sl #1}  \goodbreak \vskip.8cm} 

\def\Th#1 {
             \advance \theoremnumber by 1
             \vskip .6cm  
             \goodbreak 
             \noindent
             {\bf Theorem {\the\headlinenumber}.{\the\theoremnumber}.}
             {\sl #1}  \goodbreak \vskip.8cm}

\def\Lm#1 {
             \advance \theoremnumber by 1
             \vskip .6cm  
             \goodbreak 
             \noindent
             {\bf Lemma {\the\headlinenumber}.{\the\theoremnumber}.}
             {\sl #1}  \goodbreak \vskip.8cm}

\def\Cor#1 {
             \advance \theoremnumber by 1
             \vskip .6cm  
             \goodbreak 
             \noindent
             {\bf Corollary {\the\headlinenumber}.{\the\theoremnumber}.}
             {\sl #1}  \goodbreak \vskip.8cm} 
%
%            Formelnummerierung
%
\newcount \equationnumber

\newcount \refnumber

\def\[]    {\global 
            \advance \refnumber by 1
            [{\the\refnumber}]}

\def\# #1  {\global 
            \advance \equationnumber by 1
            $$ #1 \eqno ({\the\equationnumber}) $$ }

\def\% #1 { \global
            \advance \equationnumber by 1
            $$ \displaylines{ #1 \hfill \llap ({\the\equationnumber}) \cr}$$} 

\def\& #1 { \global
            \advance \equationnumber by 1
            $$ \eqalignno{ #1 & ({\the\equationnumber}) \cr}$$}
%
%                             Referencen
%                                
\newcount \Refnumber

\def\Ref #1 #2 #3 #4 #5 #6  {\ninerm \global
                             \advance \Refnumber by 1
                             {\ninerm #1,} 
                             {\ninesl #2,} 
                             {\ninerm #3.} 
                             {\ninebf #4,} 
                             {\ninerm #5,} 
                             {\ninerm (#6)}\nobreak} 
\def\Bookk #1 #2 #3 #4       {\ninerm \global
                             \advance \Refnumber by 1
                             {\ninerm #1,}
                             {\ninesl #2,} 
                             {\ninerm #3,} 
                             {(#4)}}
\def\Book{\cr
{\the\Refnumber} &
\Bookk}
\def\Reff{\cr
{\the\Refnumber} &
\Ref}
\def\REF #1 #2 #3 #4 #5 #6 #7   {{\sevenbf [#1]}  & \hskip -9.5cm \vtop {
                                {\sevenrm #2,} 
                                {\sevensl #3,} 
                                {\sevenrm #4} 
                                {\sevenbf #5,} 
                                {\sevenrm #6} 
                                {\sevenrm (#7)}}\cr}
\def\BOOK #1 #2 #3 #4  #5   {{\sevenbf [#1]}  & \hskip -9.5cm \vtop {
                             {\sevenrm #2,}
                             {\sevensl #3,} 
                             {\sevenrm #4,} 
                             {\sevenrm #5.}}\cr}
\def\HEP #1 #2 #3 #4     {{\sevenbf [#1]}  & \hskip -9.5cm \vtop {
                             {\sevenrm #2,}
                             {\sevensl #3,} 
                             {\sevenrm #4.}}\cr}
%
%                               Beweis
%
\def\bull{$\sqcup \kern -0.645em \sqcap$}
%
%               Definition, Remark, Example, Proof  
%
\def\Def#1{\vskip .3cm \goodbreak \noindent
                                     {\bf Definition.} #1 \goodbreak \vskip.4cm}
\def\Rem#1{\vskip .4cm \goodbreak \noindent
                                     {\it Remark.} #1 \goodbreak \vskip.5cm }

\def\Pr#1{\goodbreak \noindent {\it Proof.} #1 \hfill \bull  \goodbreak \vskip.5cm}

%
%                               Abstand
%
\def\*{\vskip 1.0cm}      

%
%            †berschriften
%
\newcount \ssubheadlinenumber

\def\SSHL #1 {\goodbreak
            \cleartheoremnumber
            \vskip 1cm
            \advance \ssubheadlinenumber by 1
{\rm \noindent {\the\headlinenumber}.{\the\subheadlinenumber}.{\the\ssubheadlinenumber}. #1}
            \nobreak \vskip.8cm \rm \noindent}
\newcount \subheadlinenumber
\def\clearsubheadlinenumber{\subheadlinenumber = 0 \relax}
\def\SHL #1 {\goodbreak
            \cleartheoremnumber
            \vskip 1cm
            \advance \subheadlinenumber by 1
            {\rm \noindent {\the\headlinenumber}.{\the\subheadlinenumber}. #1}
            \nobreak \vskip.8cm \rm \noindent}
\newcount \headlinenumber

\newcount \headlinesubnumber
\def\clearheadlinesubnumber{\headlinesubnumber = 0 \relax}
\def\Hl #1 {\goodbreak
            \cleartheoremnumber
            \clearheadlinesubnumber
            \clearsubheadlinenumber
            \advance \headlinenumber by 1
            {\bf \noindent {\the\headlinenumber}. #1}
            \nobreak \vskip.4cm \rm \noindent}

\font\twentyrm=cmr17
\font\fourteenrm=cmr10 at 14pt
\font\sevensl=cmsl10 at 7pt
\font\sevenit=cmti7 

\font\css=cmss10
\font\Rosch=cmr10 at 9.85pt
\font\Cosch=cmss12 at 9.5pt
\font\rosch=cmr10 at 7.00pt
\font\cosch=cmss12 at 7.00pt
\font\nosch=cmr10 at 7.00pt
%
%
%
%
%
%
%
%                   Symbols
%
%
%
%                              
%
%
%
%               reelle, nat\"urliche, ganze, komplexe Zahlen
%
%    Beispiel:  ${\bf E}_\alpha {s \atop {\raise 4pt \hbox{$\to$}}} 1$
%
\def\Z                 {\hbox{{\css Z}  \kern -1.1em {\css Z} \kern -.2em }}
\def\R                 {\hbox{\raise .03ex \hbox{\Rosch I} \kern -.55em {\rm R}}}
\def\N                 {\hbox{\rm I \kern -.55em N}}
\def\C                 {\hbox{\kern .20em \raise .03ex \hbox{\Cosch I} \kern -.80em {\rm C}}}

\def\r                 {\hbox{\raise .03ex \hbox{\rosch I} \kern -.45em \hbox{\rosch R}}}
\def\n                 {\hbox{\hbox{\rosch I} \kern -.45em \hbox{\nosch N}}}
\def\c                 {\hbox{\raise .03ex \hbox{\cosch I} \kern -.70em \hbox{\rosch C}}}

\def\z                 {\hbox{\kern 0.2em {\cal z}  \kern -0.6em {\cal z} \kern -0.3em  }}
\def\1                 {\hbox{\rm \thinspace \thinspace \thinspace \thinspace
                                  \kern -.50em  l \kern -.85em 1}}
\def\unit                 {\hbox{\sevenrm \thinspace \thinspace \thinspace \thinspace
                                  \kern -.50em  l \kern -.85em 1}}
%
%
%
%
%
%                   
%                              Sonderzeichen alg. QFT
%
%
%
%                              
%
%
\def\Tr                {{\rm Tr}}
\def\A                 {{\cal A}}

\def\B                 {{\cal B}} 

\def\H                 {{\cal H}} 
 
\def\O                 {{\cal O}}

\def\bra{\langle}
\def\ket{\rangle}

%
%                                
%
%
%                             Referencen
%                                
%
%

%\Z  
%\R  
%\N  
%\C  
%\Ro 
%\No 
%\Zo 
%\Co 
%\r  
%\n  
%\c  
%\ro 
%\no 
%\co 
%\z  
%\1  
%\unit  

\def\versuch #1 #2 {
\vskip -.1 cm
\global \advance \equationnumber by 1
            $$\displaylines{ \rlap{ #1 } \hfill #2  \hfill \llap{({\the\equationnumber})} } $$ 
\vskip  .1cm
\noindent}

\nopagenumbers
\def\Draft  {\hbox{Preprint \today}}
\def\firstheadline{\hss \hfill  \Draft  \hss} 
\headline={
\ifnum\pageno=1 \firstheadline
\else 
\ifodd\pageno \rightheadline 
\else \leftheadline \fi \fi}
\def\rightheadline{\sevenrm NUCLEARITY AND SPLIT FOR THERMAL QUANTUM FIELD THEORIES  
\hfill \folio } 
\def\leftheadline{\sevenrm \folio \hfill CHRISTIAN D.\ J\"AKEL}
\voffset=2\baselineskip
\magnification=1200
%**************************************************************************************************
%
%
%                  TITELSEITE
%
%
%
%**************************************************************************************************

\vskip 1cm

\noindent
{\twentyrm Nuclearity and Split for Thermal}
                  
\vskip .2cm
\noindent
{\twentyrm Quantum Field Theories}
                  
\vskip 1cm
                  
\noindent
{\it Dedicated to Prof.\ Walter Thirring on the occasion of his $70th$ birthday\footnote{$^{a)}$}
{\sevenrm The author apologizes for the substantial delay of the present version of
this article.}}

\vskip .5cm
\noindent
{\sevenrm CHRISTIAN D.\ J\"AKEL}

\noindent
{\sevenit  Dipartimento di Matematica,
via della Ricerca Scientifica, Universit\`a di Roma ``Tor Vergata'', 
I-00133~Roma, e-mail: christian.jaekel@uibk.ac.at}

\vskip .5cm     
\noindent {\sevenbf Abstract}. {\sevenrm We review the heuristic arguments
suggesting that any thermal quantum field theory, which can be interpreted as a 
quantum statistical mechanics of (interacting) relativistic particles,
obeys certain restrictions on its number of local degrees of freedom. As in the
vacuum representation, these restrictions can be expressed by a 
`nuclearity condition'. If a model satisfies this nuclearity condition,
then the net of von Neumann algebras representing the local 
observables in the thermal representation has the split property.}

\vskip 1 cm

%**************************************************************************************************
%
%
%                  Einleitung
%
%
%
%**************************************************************************************************

\Hl{Introduction}

\noindent
Haag and Swieca [HS] suggested that a quantum field theory, which
allows a particle interpretation, should have specific phase--space properties
in the vacuum sector. 
This idea motivated Buchholz and Wichmann [BW] to investigate the restrictions on the 
energy level density in the vacuum sector
imposed by the existence of thermal equilibrium states. The result of their careful analysis is 
a `nuclearity condition' which on one hand is satisfied in all models of physical 
relevance and on the other hand tightens up the axiomatic structure considerably.
Numerous results in algebraic QFT (e.g., the existence 
of KMS states [BJ b], a local version of the Noether theorem [BDL], etc.)  
emerged from this refinement of the axiomatic structure.

In this article we formulate a nuclearity condition for thermal field
theories (TFTs) and investigate its consequences in the axiomatic framework.   
Thermal representations are always reducible. Therefore their structural properties 
are somehow complementary to the ones known from zero temperature quantum field theory.
Nevertheless a number of basic physical properties like the Reeh--Schlieder 
property, the Schlieder property and the Borchers property hold; in fact, they can be 
established without taking recourse 
to results from the vacuum sector [J\"a a,b]. 
What is known sofar  
concerning the statistical independence of local observables can be summarized as follows: 

\Th{Assume a TFT is specified 
by a net
\# { \O \to {\cal R}_\beta (\O), \qquad \O \subset \R^4,}
of von Neumann algebras, subject to the standard assumptions stated explicitly
in the next section (see p.6). Now let $\O $, $\hat{\O}$ denote 
a pair of space--time regions in Minkowski space such that the closure of the open (not necessarily  
bounded) region~$\O$ is contained the interior of~$\hat{\O}$. (This geometrical situation will
be denoted by $\O \subset \subset \hat{\O}$ in the sequel.)  
It follows that  
\vskip .3cm
\halign{ #  \hfil & \vtop { \parindent =0pt \hsize=36,6em
                            \strut # \strut} \cr 
(i)  & for every normal state $\omega_1$
on ${\cal R}_\beta (\O)$ and every normal state $\omega_2$ on 
${\cal R}_\beta (\hat{\O})' $ 
there exists a normal 
state $\omega$ on $\B(\H_\beta)$ 
such that 
\# { \omega_{| {\cal R}_\beta (\O)} = \omega_1 \qquad \hbox{and} \qquad 
\omega_{| {\cal R}_\beta (\hat{\O})' } = \omega_2 .}
\cr
(ii)   & for every state $\phi_1$
on ${\cal R}_\beta (\O)$ and every state $\phi_2$ on ${\cal R}_\beta (\hat{\O})' $ there 
exists a state $\phi$ on $\B(\H_\beta)$ 
such that 
\# { \phi (AB) = \phi_1 (A) \phi_2 (B) }
for all $A \in {\cal R}_\beta (\O)$ and all $B \in {\cal R}_\beta (\hat{\O})' $.
\cr}  
\vskip .2cm
\noindent
As usual, the Hermitian elements of ${\cal R}_\beta (\O)$ are interpreted as 
the observables which can be measured at times and locations in $\O$.}

\Rem{As it turned out, the two statements (i) and (ii) are equivalent [FS]. As we will
show in Section~5, there remain only two possibilities (see also [Bu][Su][FS]):
\vskip .2cm
\halign{ #  \hfil & \vtop { \parindent =0pt \hsize=36,6em
                            \strut # \strut} \cr 
(i)  & if there exists at least one
normal state~$\phi$ on $\B(\H_\beta)$, which is a product state for 
${\cal R}_\beta (\O)$ and ${\cal R}_\beta (\hat{\O})'$, then there exist sufficiently many.
More precisely, there exists, 
for any pair of normal states $\omega_1$ of ${\cal R}_\beta (\O)$ and
$\omega_2$ of ${\cal R}_\beta (\hat{\O})'$, a normal
state~$\omega_{1,2}$ on~$\B(\H_\beta)$, which is a normal extension of 
$\omega_1$ and~$\omega_2$ and a product state for ${\cal R}_\beta (\O)$ 
and ${\cal R}_\beta (\hat{\O})'$. 
The existence of these normal product states is equivalent to the existence of 
a type~I factor ${\cal N}_\beta$ such that 
\# {{\cal R}_\beta (\O) \subset {\cal N}_\beta \subset {\cal R}_\beta (\hat{\O}) ;}
in this case the 
inclusion ${\cal R}_\beta (\O) \subset {\cal R}_\beta (\hat{\O})$ is called 
split. (For a general discussion of split inclusions see [DL].) 
\cr
(ii)   & all normal partial states  
have normal extensions, none of which is a product state,
and also all partial states have extensions to product states, 
none of which is normal.
\cr}  
}

Just as in the vacuum sector, the missing piece of information in order to favour one of the
two possibilities (i) or (ii) stated in the previous Remark is encoded in
the phase--space properties of a given TFT:  
In Section 4 we prove that the split property (4) 
can be derived from an appropriate  nuclearity condition, which we expect to be
satisfied in all physically relevant TFTs; thus we can rule out 
possibility~(ii) for those theories.  
In Section 3 we give a self-contained, heuristic derivation
of the nuclearity condition in the thermal sector, based on the work of several 
authors (see for instance [BW][BD'AL b][BY], etc.). 
Section~5 exploits several equivalent 
formulations of the split property; whereas in the final section we list some 
of its implications.

\vskip 1cm

\goodbreak

\Hl{Preliminary Definitions and Results}

\noindent
Let us briefly recall the standard setup:
In the Araki--Haag--Kastler framework [H] a quantum field theory (QFT) is
specified by a net 
\# { \O \to \A (\O) , \qquad \O \subset \R^4, } 
of $C^*$-algebras.
$\A (\O)$ represents the algebra generated by the observables which can be measured 
in the space--time region $\O$. 

\SHL{Representation Independent Properties}

\noindent
The net $\O \to \A(\O)$ has certain
properties irrespective of the (global) properties of the (inital) physical 
state under consideration:

\vskip .2cm 
\noindent
i.)  The net $\O \to \A(\O)$ is isotonous, i.e., there exists a unital embedding
\# {\A(\O_1) \hookrightarrow \A(\O_2)
\qquad \hbox{if} \quad \O_1 \subset \O_2.} 
Isotony allows us to consider the quasi-local algebra
\# {\A:= \overline{ \cup_{ {\cal O}  \subset \r^4} \A(\O) }^{\, C^*},}
which is defined as the $C^*$-inductive limit of the local algebras.
The elements of $\A$ are called quasi-local observables; they can be approximated in norm 
topology by strictly local elements; the total energy, total charge, etc., 
are considered to be  unobservable; these quantities refer to infinitely extended regions
and can not be controlled by local measurements.

\vskip .2cm 
\noindent
ii.)  Observables localized in spacelike separated space--time regions commute:
\#
{\A (\O_1) \subset \A^c ( \O_2) \quad \hbox{\rm if} \quad \O_1 \subset \O_2'.}
Here $\O '$ denotes the spacelike complement of $\O$ and 
$\A^c (\O)$ denotes the set of operators in~$\A$ which commute with all operators in $\A(\O)$.

\vskip .2cm 
\noindent
iii.) The space--time symmetry of Minkowski space manifests itself
in the existence of a representation 
\# {\alpha \colon ( \Lambda, x) \mapsto \alpha_{ \Lambda, x} \in Aut (\A) ,
\qquad (\Lambda, x) \in {\cal P}_+^\uparrow}, 
of the (orthochronous)
Poincar\'e group ${\cal P}_+^\uparrow$. Lorentz transformations $\Lambda$  and space--time 
translations~$x$ act geometrically:
\# {\alpha_{ \Lambda, x} \bigl( \A (\O) \bigr) 
= \A (\Lambda \O + x)  \qquad \forall (\Lambda, x) \in {\cal P}_+^\uparrow.} 

\Rem{Without loss of generality, we may assume that the space--time translations
$\alpha \colon \R^4 \to Aut (\A)$ are strongly continuous. In this case 
the energy--momentum transfer of an element $a \in \A$ has a representation independent
meaning. One can 
define the Fourier transforms of the operator valued functions 
$x \mapsto \alpha_x  (a) $, $a \in \A$, in the sense of distributions: 
for each $f \in L^1 (\R^4, {\rm d}^4x)$ the expression
\# { \alpha_f  (a) := \int {\rm d}^4 x \, f(x) \alpha_x  (a) ,   \qquad  a \in \A,} 
exists as a Bochner integral in $\A$, since $ \| \alpha_f  (a) \| \le \|f \|_1  \| a \|$.
The {\it energy--momentum  transfer} of an element $a \in \A$ is defined as the 
smallest closed subset 
$\tilde \O \subset \R^4$ such that 
\# { \alpha_f  (a) = 0    \qquad  \forall f \in L^1 (\R^4) \quad \hbox{with} \quad
\hbox{supp} \, \tilde f \subset \R^4 \setminus \tilde \O,} 
where $\tilde f$ denotes the Fourier transform of $f$ (cf.\ [BV]).} 

\Rem{For the present article we may restrict our attention to
the (strongly continuous) one-parameter subgroup of time translations 
$\tau \colon \R \to Aut(\A)$. Of course, it acts geometrically, i.e.,
\# {\tau_t \bigl( \A (\O) \bigr) 
= \A (\O +te)  \qquad \forall t \in \R.} 
Here $e$ is a unit vector denoting the time 
direction with respect to a given Lorentz frame.} 

\SHL{Representation Dependent Properties}

\noindent
The relevant states describing thermal equilibrium 
are distinguished within the set of all time invariant normalized, positive linear
functionals of~$\A$ by their stability properties with respect to timelike translations.
They are conveniently characterised by the KMS condition [HHW]:

\Def{A state
$\omega_\beta$ over $\A$ is called a $(\tau , \beta )$-KMS state for 
some $\beta \in \R \cup \{ \pm \infty \}$, if 
\# { \omega_\beta \bigl( a \tau_{i \beta} (b) \bigr) = \omega_\beta (b a) }
for all $a, b$ in a norm dense,
$\tau$-invariant $*$-subalgebra of $\A_\tau$. (${\A}_{\tau} \subset \A$
denotes the set of analytic elements for~$\tau$.)}

Given a KMS state $\omega_\beta$, the GNS construction gives rise to a 
Hilbert space~$\H_\beta$ and a 
representation $\pi_\beta$, called a thermal representation, of $\A$.
The algebra ${\cal R}_\beta := \pi_\beta (\A)''$ possesses a cyclic 
(due to the GNS construction) and separating (due to the KMS condition) 
vector~$\Omega_\beta$ such that
\# { \omega_\beta (a) = \bigl( \Omega_\beta \, , \, \pi_\beta (a) \Omega_\beta \bigr) 
\qquad \forall a \in \A.}

\Not{The state vector $\Omega_\beta$ induces a natural extension of $\omega_\beta$ to $\B(\H_\beta)$.
By abuse of notation the same symbol, namely $\omega_\beta$, will be used to denote both 
the extension and the original state.} 

A KMS state is time invariant. Therefore
the one-parameter group of unitaries implementing the
time translations $\tau \colon  \R \to Aut (\A)$ in the 
representation~$\pi_\beta$ is uniquely specified by putting 
\# { {\rm e}^{i H_\beta t} \pi_\beta (a) \Omega_\beta 
:= \pi_\beta \bigl( \tau_t (a) \bigr)  \Omega_\beta 
\qquad \forall
a \in \A .}

\Rem{In order to support our heuristic argumentation later on, let us assume, 
just for a moment, that
the time evolution $\tau$ is inner, i.e.,  $\tau$ is generated by an 
element~$h$ of the $C^*$-algebra~$\A$:
\# {\tau_t (a) = {\rm e}^{iht} a {\rm e}^{-iht} \qquad \forall a \in \A . }
It follows that the generator $H_\beta$ can be identified as  
\# { H_\beta  =  \pi_\beta (h) - J_\beta \pi_\beta (h) J_\beta ,}
where $J_\beta$ is be the modular conjugation associated with the pair 
$\bigl( {\cal R}_\beta, \Omega_\beta \bigr)$. Note that even in this case $H_\beta$ 
and $\pi_\beta (h)$ differ from each other
not only by the thermal expectation value of the energy $\omega_\beta (h)$, but 
in the removal of an operator of ${\cal R}_\beta '$. If one withdraws the 
(spatial or/and momentum) cut-offs which are
implicitly enforced by requiring that $h \in \A$, then 
the decomposition (18) of $H_\beta$ is no longer possible.}  

We will $\underline {\hbox{not}}$ require that spacelike translations
can be unitarily implemented in the representation $\pi_\beta$, 
since spatial translation invariance may be spontaneously broken in a KMS state.

\vskip .5cm

We could now continue to derive more specific properties of
the net 
\# { \O \to {\cal R}_\beta (\O) := \pi_\beta \bigl(\A(\O)\bigr)'' } 
from first principles\footnote{$^\dagger$}{\sevenrm
For instance, if the KMS state $\scriptstyle \omega$ is extremal and the time evolution 
is asymptotically abelian, i.e.,
$\scriptstyle \lim_{t \to \infty} \| [ a, \tau_t (b) \| = 0 $ 
for all $\scriptstyle a,b \in \A$, 
then $\scriptstyle \Omega_\beta$ is the unique --- up to a phase --- time invariant vector 
in $\scriptstyle \H_\beta$.}, but we rather prefer to conclude our outline of the general setting
at this point; a more detailed description will be presented elsewhere. Instead we 
emphasize that in the rigorous part of this article, which starts in Section~4,
we will exclusively rely on the following 

\vskip 1cm

\noindent
{\bf Standard Assumptions of Thermal Field Theory.} A TFT is specified by 
a von Neumann algebra ${\cal R}_\beta$ with a cyclic and separating vector
$\Omega_\beta$ together with a net of subalgebras 
\# { \O \to {\cal R}_\beta (\O),  } 
which is subject to the following conditions: 
\vskip .3cm
\halign{ \indent #  \hfil & \vtop { \parindent =0pt \hsize=32em
                            \strut # \strut} \cr 
i.)     & the subalgebras associated with spacelike 
separated space--time regions commute, i.e.,
\# { {\cal R}_\beta (\O_1) \subset {\cal R}_\beta (\O_2)'
\qquad \hbox{if} \qquad \O_1 \subset \O_2'.}
\cr
ii.)     & the modular group $t \mapsto \Delta^{it}$ associated
with the pair $({\cal R}_\beta, \Omega_\beta)$ coincides --- up to rescaling --- with the  
time evolution and therefore acts geometrically, i.e.,
\# {  {\rm e}^{i H_\beta t} {\cal R}_\beta (\O)   {\rm e}^{-i H_\beta t} 
= {\cal R}_\beta (\O + t e) 
\qquad 
\forall t \in \R.}
Here $e$ is the unit vector denoting the time 
direction w.r.t.\  the distinguished  rest frame and the modular operator 
$\Delta_\beta = \exp (- \beta H_\beta)$. 
\cr
iii.)     & $\H_\beta$ is separable and $\Omega_\beta$ is the unique 
           --- up to a phase --- time invariant vector in $\H_\beta$. 
\cr
iv.)    &  $\Omega_\beta$ is cyclic for the local algebra ${\cal R}_\beta (\O)$,
           where $\O$ is an open subset of~$\R^4$;
           i.e., 
\# { \overline { {\cal R}_\beta (\O) \Omega_\beta} = \H_\beta .}
This property is called the {\it Reeh--Schlieder property}. 
\cr}

\Rem{The Reeh--Schlieder property follows 
from the relativistic KMS condition of Bros and Buchholz [BB] provided the net $
\O \to {\cal R}_\beta (\O)$
satisfies additivity~[J\"a~b]. As has been shown by 
Junglas [Ju] the Reeh--Schlieder 
property can as well be derived from the 
standard KMS condition, as long as  $ \omega_\beta$ is locally normal 
w.r.t.\ the vacuum representation.}
\vskip 1cm

\Hl{The Nuclearity Condition in the Thermal Sector}

\noindent
In quantum mechanics the number of states in a finite phase--space
volume is finite (it is of the order ${ \hbox{\it phase--space   
volume} / 2 \pi \hbar}$). For QFTs the situation is --- 
due to imperfect localization properties  ---
more delicate. Here we claim that even for thermal field theories the set of normal states 
representing excitations which are `well-localized in phase--space'
is `small' (although not finite dimensional). More precisely,
we propose to use --- for $\lambda > 0$ and $\O$ bounded ---
\# { {\cal S}_\beta (\O, \lambda) 
= \{ {\rm e}^{- \lambda | H_\beta |} A \Omega_\beta \in \H_\beta: A \in {\cal R}_{\beta} (\O),
\| A \| \le 1 \} } 
as an appropriate set of (not normalized) state vectors describing excitations 
of the KMS state which 
are well localized both in momentum and coordinate space. 
(Recall that all normal states are vector states in a thermal
representation. Therefore ${\cal S}_\beta (\O, \lambda)$ specifies a set of normal states.) 
It is the aim of the following two subsections to make precise what we mean by claiming 
that these normal states are well-localized in phase--space and in which sense 
${\cal S}_\beta (\O, \lambda)$ is small.

\SHL{Excitations of a Thermal State}

\noindent
A normal state will be called a {\it strictly localized 
excitation} of the KMS state, if it can not 
be distinguished from the thermal equilibrium 
by measurements in the spacelike complement~$\O'$ of $\O$. Identifying state vectors
and normal states, the strictly localized excitations can be described by the 
following set of vectors: 
\# {{\cal L}_\beta (\O) := \{ \Psi \in \H_\beta : (\Psi , B \Psi) = \omega_\beta (B)
\quad \forall B \in {\cal R}_\beta (\O') \} \subset \H_\beta . }
Strict localization is a rather cumbersome notion: in general, 
not even linear combinations of elements of ${\cal L}_\beta (\O)$ will 
belong to ${\cal L}_\beta (\O)$. 
This problem can be circumvented by relaxing the localization criterion:
for extremal KMS states  
decent infrared properties of~$H_\beta$ --- as specified in (30) below ---
ensure that 
we can as well use  
\# {{\cal S}_\beta (\O) := \{ A \Omega_\beta \in \H_\beta: A \in {\cal R}_{\beta} (\O),
\| A \| \le 1 \} } 
as a suitable set of state vectors with good localization properties in coordinate space. 
This argumentation is supported by the cluster theorem for KMS states presented below [J\"a~c]. 

\Not{The state vector $A\Omega_\beta$, $A \in {\cal R}_\beta (\O)$, induces a 
state $\omega^A_\beta$, specified by  
\# {{\cal R}_\beta  \ni C \mapsto \omega^A_\beta (C) 
:= {(A \Omega_\beta \, , \, C A \Omega_\beta ) \over \| A \Omega_\beta \|^2} .} 
Since the KMS state distinguishes a restframe, there exists a distinguished time direction 
$e = (1,0,0,0)$.
Let $e_\bot = (0, r, s, t)$, $r,s,t \in \R$, $\|e_\bot \| = 1$, be a spatial vector 
w.r.t.\ the distinguished restframe. Finally, consider the double cone 
\# { \O := (V_+ - \lambda e) \cap (V_- + \lambda e), \qquad \lambda \in \R^+,}
where the forward (resp.\ backward) light cone is 
$V_\pm = \{ x \in \R^4 : x^0 {> \atop <} \pm | \vec x | \}$.} 

\Prop{Let $\O$ be the double cone introduced in (28). Furthermore,
let $A\in {\cal R}_\beta (\O)$, let $\delta >0$ be a real number
and let $B \in {\cal R}_\beta (\O + \delta e_\bot)$. 
Assume there exist 
positive constants $m>0$ and $C ({\cal O})$ such that   
\# { \bigl\|  {\rm e}^{- {\lambda \over 2} | H_\beta|} \bigl( A - 
\omega_\beta ( A ) \bigr) 
\Omega_\beta  \bigr\|  
\le C ({\cal O})  \lambda^{-m}  \, \| A \|  . }
It follows that (for $\delta$ large compared to $\beta$ and the diameter of $\O$) 
the expectation values in the
state $\omega^A_\beta$
converge to the thermal expectation values as the
spacelike distance $\delta$ of the regions of $\O$ and 
$\O + \delta e_\bot$ increases:
\# { \Bigl| \omega^A_\beta (B) 
- \omega_\beta (B)  \Bigr| \le  
const. \,  \delta^{- 2m}
 { \| A \|^2 \over \| A \Omega_\beta \|^2}  \, \| B \|  . }
(The $const. \in \R^+$ is independent of $\delta$, $A$ and $B$.) }

\Rem{Due to the Reeh--Schlieder property ${\cal S}_\beta (\O ) $
is dense in $\H_\beta$. But in order to recognize the deviations from the thermal 
expectation values in the region $\O + \delta e_\bot$ (whose spacelike 
distance to~$\O$ --- neglecting the diameter of $\O$ --- may be
only several times the thermal wavelength), it is necessary to increase
the ratio between `cost and effort' $\| A \| 
/ \| A \Omega_\beta \|$ on the r.h.s.\ of (30) or the sensitivity of the measurement,
i.e., the norm of $B \in {\cal R}_\beta (\O + \delta e_\bot)$.
Thus the essential point in the definition of ${\cal S}_\beta (\O ) $ is that 
the requirement $\| A \| \le 1$. It implies that a vector $A\Omega_\beta$,
which describes an excitation that is not essentially localized in $\O$,
has  a rather small norm.}

In order to specify normal states, which are also
well-localized in momentum space, it is sufficient to 
restrict the energy transferred by 
the element~$A \in {\cal R}_\beta (\O)$ onto the KMS state. 
(As we have pointed out 
the energy momentum transfer has a representation independent meaning.)
This can be achieved by 
taking time averages 
\# {  { 1 \over \sqrt { 2 \pi }}
\int {\rm d}t \, f(t) {\rm e}^{i H_\beta t} A \Omega_\beta 
= \tilde f(H_\beta) A  \Omega_\beta ,
\qquad A \in {\cal R}_\beta (\O),}
with suitable testfunctions~$f(t)$, 
whose Fourier transforms~$\tilde f (\nu)$ decrease exponentially [BD'AL b].
A convenient choice is $\tilde f (\nu) := {\rm e}^{- \lambda | \nu |}$ with $\lambda >0$. 
We conclude that if  $\lambda > 0$, then the elements of 
\# { {\cal S}_\beta (\O , \lambda) 
= \{ {\rm e}^{- \lambda | H_\beta |} A \Omega_\beta \in \H_\beta: A \in {\cal R}_{\beta},
\| A \| \le 1 \} } 
induce vector states with good localization properties in coordinate and
momentum space.

\SHL{Finite Volume Gibbs States}

\noindent
Let us, for simplicity, consider massive particles in a finite volume $V$
in the grand canonical ensemble.  
The energy spectrum will then be discrete and the theory can be conveniently
described in terms of energy eigenfunctions $\Psi_i$ in Fockspace $\H_{\cal F}$:
\# { H_{\cal F}  \Psi_i = E_i \Psi_i \qquad \hbox{with} 
\qquad E_i \in \R^+ \cup \{ 0 \}.}
$H_{\cal F} \ge 0$ denotes the Hamilton operator acting on a dense domain in $\H_{\cal F}$.
The grand canonical equilibrium state (at zero chemical potential)
is described by a density matrix $\rho_\beta \in \B(\H_{\cal F})$:
\# { \A \ni a \mapsto   \Tr \, \rho_\beta \, \pi_{\cal F} (a) . }
Here $\pi_{\cal F} (a)$ denotes the Fock space representation of an element $a \in \A$.
For a given inverse temperature $\beta$ the grand canonical equilibrium state is
unique, once the boundary conditions for the Hamiltonian $H_{\cal F}$
are fixed. As long as the volume $V$ of the `box' is finite, it is reasonable to assume that
${\rm e}^{- \beta H_{\cal F}}$ is traceclass. In this case
the Gibbs density matrix is just
\# { \rho_\beta = { {\rm e}^{- \beta H_{\cal F}} \over \Tr \, {\rm e}^{- \beta H_{\cal F}} },
\qquad \beta > 0.}
However,
as the volume $V$ of the box increases, the spacing of the eigenvalues decreases
drastically. In the thermodynamic limit the spectrum of the Hamiltonian 
becomes continuous and ${\rm e}^{- \beta H_{\cal F}}$ can no longer be traceclass. 
In order to characterize the phase--space properties of an
infinite system it is therefore necessary to look for more decent properties, 
which may survive the thermodynamic limit. We start with a rather general
classification:

\Def{A continuous linear mapping 
$\Theta$ from a Banach space~${\cal E}$
to another Banach space~${\cal F}$ is said to be of 
type~$l^p$, $p > 0$, if there exists
a sequence of linear mappings $\Theta_k$ of rank $k$ such that
\# { \sum_{k=0}^\infty \| \Theta - \Theta_k \|^p < \infty  .}  
$\Theta$ is said to be nuclear, if 
$\Theta$ is of type~$l^p$ for $p =1$. 
$\Theta$ is said to be of type  $s$, if 
$\Theta$ is of type~$l^p$ for all $p > 0$. 
The order $q$ of the map  
$\Theta$ is defined as the nonnegative number 
(if it exists)
\# { q = \limsup_{\epsilon \searrow 0} {\ln \ln N( \epsilon ) \over \ln 1 /\epsilon },}
where $N(\epsilon)$, the $\epsilon$-content of $\Theta$, is the maximal number of
elements $E_i$ in the unit ball of~${\cal E}$ such that $\bigl\| \Theta (E_i - E_k) \bigr\| > 
\epsilon$ if $i \ne k$.}

\Rem{The maps of fixed type form an ideal in the space of all bounded maps
between Banach spaces [P].}

A delicate part of the argument concerns the relation between the Fock representation and the
GNS representation induced by
\# { \rho_\beta = { 1 \over  \sum_k {\rm e}^{- \beta E_k} }
\sum_ i   {\rm e}^{- \beta E_i}  
\, \, | \Psi_i \ket \bra \Psi_i | ,
\qquad \beta > 0.}
The GNS representation is unitarily equivalent to the representation  $\pi_{\beta, V}
\colon a \mapsto \pi_{\beta, V} (a) := \pi_{\cal F} (a ) \otimes \1 $
constructed with the cyclic vector
\# { |  \sqrt {\rho_\beta } \ket =  { 1 \over  \sqrt{ \sum_k  {\rm e}^{- \beta E_k } } } 
\sum_i    {\rm e}^{- \beta E_i /2  }  \, \,   \Psi_i \otimes \Psi_i
\in \H_{\cal F} \otimes \H_{\cal F} . }
One finds
\# { J_{\beta, V} \, \pi_{\beta, V} (a) \, J_{\beta, V} = \1 \otimes \pi_{\cal F} (a ) 
\qquad \forall a \in \A.} 
Here $J_{\beta, V}$ denotes the modular conjugation associated with
the pair $\bigl( \pi_{\beta, V} (\A)'', |  \sqrt {\rho_\beta } \ket \bigr)$.
Corroborating the insights gotten from inner time evolutions (18) we conclude that
in the representation  $\pi_{\beta, V}$ the time evolution is generated by 
\# { H_{\cal F} \otimes  \1  - \1 \otimes H_{\cal F} .}
We can now estimate

\SHL{The Size of ${\cal S}_\beta (\O, \lambda)$}

\noindent
Let us consider the map $\Theta_V \colon \pi_{\beta, V} (\A (\O)) \to 
\H_{\cal F} \otimes \H_{\cal F}$, 
\# { A \mapsto \exp \Bigl( - \lambda \bigl| H_{\cal F} \otimes  \1  
- \1 \otimes H_{\cal F} \bigr| \Bigr) \, \, 
( A \otimes \1  ) \, \, |  \sqrt {\rho_\beta } \ket .}
A straight forward computation yields
\& { \Theta_V (A) & =
\sum_{i,j} {\rm e}^{- \lambda | E_i - E_j | } ( A_{i,j} \otimes \1 )
{   {\rm e}^{- \beta E_j /2  } 
\over \sqrt{ \sum_k  {\rm e}^{- \beta E_k } } } \, \,   \Psi_j \otimes \Psi_j
\cr
& = {1 \over \sqrt{ \sum_k  {\rm e}^{- \beta E_k } }}
\sum_{i,j} {\rm e}^{- \lambda | E_i - E_j | - \beta E_j /2  }
( A_{i,j} \Psi_j \otimes \Psi_j) } 
where
\# { A_{i,j} := | \Psi_i \ket \bra \Psi_i | A 
| \Psi_j \ket \bra \Psi_j |}
is a rank $1$ operator. Moreover, the sum in (43) is convergent for $\lambda >0$;
thus $\Theta_V$ is a nuclear map;
in fact since $\Theta_V$ is nuclear for all $\lambda > 0$  it is even 
an element of all Schatten--von Neumann classes, thus it is of type $s$ (order 0).

As long as long--range correlations play no significant role\footnote{$^\dagger$}{\sevenrm 
If long-range correlations are not negligible, 
then boundary effects may spoil this part of our argument. But as long as 
$\scriptstyle H_\beta$ has decent infrared properties, the cluster theorem indicates that
this should not be the case.}, we may compare
the theory in a large (compared to the spatial 
extension of the bounded space--time region $\O \subset \R^4$) but finite volume $V$ with 
the infinite volume theory.
Disregarding boundary effects, there should exist a similarity transformation
(i.e., a bounded, invertible map) $S$ from the finite volume Hilbert space 
\# { \H_{\cal F} \otimes \H_{\cal F} = 
\overline{ \pi_{\beta, V} (\A)''  |  \sqrt {\rho_\beta } \ket } }
onto $\H_\beta$ such that
\# { {\rm e}^{- \lambda | H_\beta |} \pi_\beta \bigl(\A_1 (\O) \bigr) \Omega_\beta 
\subset  S \cdot \exp \Bigl( - \lambda \bigl| H_{\cal F} \otimes  \1  
- \1 \otimes H_{\cal F} \bigr| \Bigr) \, \, 
( \pi_{\cal F} (\A_1 (\O)) \otimes \1  ) \, \, |  \sqrt {\rho_\beta } \ket  ,}
where $\A_1 (\O)$ denotes the unit ball in $\A (\O)$. 
If this is the case, then  the map 
\# { \Phi_V \colon \pi_\beta \bigl(\A (\O) \bigr) \to \H_{\cal F} }
specified by 
\# { \Phi_V (A) = \exp \Bigl( - \lambda \bigl| H_{\cal F} \otimes  \1  
- \1 \otimes H_{\cal F} \bigr| \Bigr)
\,  S^{-1}  
{\rm e}^{- \lambda | H_\beta  |} A \Omega_\beta ,}
is bounded by $1$ if $\| A \| = 1$. Hence --- for $\O$ bounded --- 
the map 
$\Theta_{\lambda, {\cal O}} \colon 
\pi_\beta \bigl(\A (\O) \bigr) \to \H_\beta$,
\# { A \mapsto {\rm e}^{- \lambda | H_\beta |} A \Omega_\beta  }
which is obtained by composing 
$\exp \Bigl( - \lambda \bigl| H_{\cal F} \otimes  \1  - \1 \otimes H_{\cal F} \bigr| \Bigr)$ with 
the bounded maps $\Phi_V$ and~$S$, respectively, is of type $s$ (order 0) too, 
for any $\lambda > 0$ and any $\beta > 0$.

\vskip 1cm

\Hl {The Split Property in the Thermal Sector} 

\noindent
To summarize the previous section, we propose a nuclearity condition,
which should be checked in models: 
for fixed $\beta > 0$ and any bounded space--time region $\O \subset \R^4$
the maps $\Theta_{\lambda, {\cal O}}\colon {\cal R}_\beta (\O)  
\to  \H_\beta$
\# { A \mapsto  {\rm e}^{- \lambda | H_\beta |} A \Omega_\beta  }
should be of type $s$ (order 0) for any $\lambda > 0$. This condition will now 
serve as the starting point for our derivation of the split property in the thermal sector.

We start with a reformulation of this condition, which will be more convenient
in the sequel. The following (simplified) Lemma is due to Buchholz, D'Antoni and 
Longo [BD'AL b]. 
For the sake of completeness we reproduce their proof, adjusting the notation
such that it confirms with our conventions.

\Lm{If the maps $\Theta_{\lambda, {\cal O}}$ are of order $q=0$ 
for all $\lambda > 0$, then the maps 
\# { 
 A  \mapsto {\rm e}^{- \lambda H_\beta} A   \Omega_\beta , \qquad A \in {\cal R}_\beta (\O),} 
are of order $q=0$ for all $0 < \lambda < \beta / 2$.}

\Pr{Let $A \in {\cal R}_\beta (\O)$ and let 
$P^\pm$ denote the projections onto the (strictly) positive and negative 
spectrum of $H_\beta$, respectively. If the map $\Theta_{\lambda, {\cal O}}$ is of order $0$, 
then the map $ A \mapsto {\rm e}^{- \lambda H_\beta } P^+
A \Omega_\beta$
is also of order $0$, since ${\rm e}^{- \lambda H_\beta } P^+ =
 P^+ {\rm e}^{- \lambda | H_\beta| }$. The modular group $t \mapsto \Delta^{it}$
associated with the pair 
$({\cal R}_\beta, \Omega_\beta)$ coincides,
up to the rescaling $ t \mapsto -t \beta $, 
with the time evolution $t \mapsto {\rm e}^{it H_\beta }$. 
Taking advantage of the associated modular conjugation $J$ 
we find: 
\& {  {\rm e}^{ - \lambda  H_\beta } P^- A \Omega_\beta
&=  P^- {\rm e}^{-  \lambda H_\beta  }
J {\rm e}^{ - { \beta \over 2} H_\beta  }  A^* \Omega_\beta 
\cr
&=  J P^+  
{\rm e}^{- ( \beta / 2 - \lambda) H_\beta  }  A^* \Omega_\beta,
\qquad 0 \le \lambda \le \beta / 2} 
Since $ J$ is bounded,
this equality implies that the map $ A \mapsto P^-
{\rm e}^{(\beta /2 - \lambda) H_\beta }  A  \Omega_\beta$ is, for $ 0 < \lambda < \beta / 2$,
of order $0$, too. The maps of order $0$ form a linear space. 
It follows that the maps $A  \mapsto {\rm e}^{- \lambda H_\beta} A   \Omega_\beta$
are of order $0$ for the given 
range of~$\lambda$.}

Given an inclusion $\O \subset \subset \hat{\O} $ of two space--time regions, 
our task is to show that the von Neumann algebra generated by ${\cal R}_\beta (\O)$ and
${\cal R}_\beta (\hat{\O})'$ is isomorphic to the $W^*$-tensor product of the two 
algebras, i.e.,
\# { {\cal R}_\beta (\O) \vee {\cal R}_\beta (\hat{\O})' 
\cong
{\cal R}_\beta (\O) \otimes {\cal R}_\beta (\hat{\O})' .}
We will show later on that the split property (4) is a direct consequence of (53).
The first step is to insert two bounded
space--time regions $\O_1$, $\O_2$ in between $\O$ and $\hat{\O}$ such that
\# { \O \subset \subset \O_1 \subset \subset \O_2 \subset \subset \hat{\O} .}
Following Buchholz and Wichmann [BW], we consider two representations of  
\# { {\cal C}_\beta ( \O_1 , \O_2) :=
{\cal R}_\beta (\O_1) \odot {\cal R}_\beta (\O_2)', }
the algebraic tensor product of ${\cal R}_\beta (\O_1)$  
and ${\cal R}_\beta (\O_2)'$:
the first one acts on $\H_\beta$ and is given by 
\# {\pi \Bigl(  \sum_k A_k \odot B_k \Bigr) =  \sum_k A_k  B_k 
\qquad
{\rm for}
\quad
 A_k \in {\cal R}_\beta (\O_1),   
\quad
 B_k \in {\cal R}_\beta (\O_2)'.}
The operators in ${\cal R}_\beta (\O_1)$ and ${\cal R}_\beta (\O_2)'$ commute, 
so $\pi$ defines a $*$-representation of the algebraic tensor product. 
The second representation, denoted by $\pi_p$, acts on $\H_\beta \otimes \H_\beta$ 
and is determined by
\# {\pi_p \Bigr( \sum_k A_k \odot B_k \Bigl) =  \sum_k A_k \otimes B_k 
\qquad
{\rm for}
\quad
A_k \in {\cal R}_\beta ( \O_1 ),   
\quad
B_k \in {\cal R}_\beta (\O_2)'.}
As recently shown by the author [J\"a a], 
the Schlieder property holds for the pair ${\cal R}_\beta (\O_1)$ and 
${\cal R}_\beta (\O_2)'$; i.e.,
\# {AB=0 \quad \Rightarrow \quad A = 0 \quad \hbox{or} \quad B= 0 \qquad 
\forall A \in {\cal R}_\beta (\O_1), \quad  
B \in {\cal R}_\beta (\O_2)'.}
It follows that $\pi_p$ is well defined: $\sum_k  A_k \odot B_k = 0 \Rightarrow 
\sum_k A_k \otimes B_k = 0$.

\vskip .3cm
The next step is to show that the representations $\pi$ and $\pi_p$ 
of ${\cal C}_\beta ( \O_1 , \O_2)$ are not disjoint.
This follows --- up to minor adjustments --- from a result of Buchholz and Yngvason~[BuY]:

\Prop{Let $\O_1$ be a bounded space--time region and assume 
there exists some $\delta > 0$ such that $\O_1 + te \subset \O_2$ for all
$|t| < \delta$. Let ${\cal R}_\beta (\O_2)'_*$ denote the predual 
of ${\cal R}_\beta (\O_2)' $.
 It follows that the map 
$  \Xi_{ \beta, *} \colon  {\cal R}_\beta (\O_1)  
\to   {\cal R}_\beta (\O_2)'_* $  given by 
\# { A  \mapsto    ( \Omega_\beta \, ,  A \, . \, \, \Omega_\beta ) } 
is nuclear.    }

\Pr{Let $ A \in {\cal R}_\beta (\O_1)$ and   
$B \in {\cal R}_\beta (\O_2)'$. The function
\# { z \mapsto
(  \Omega_\beta \, , \,  
B  {\rm e}^{ iz  H_\beta} A \Omega_\beta) }
is analytic in the strip $0 < \Im z < \beta / 2$, while the function
\# { z \mapsto
( {\rm e}^{  i \bar z H_\beta} A^* \Omega_\beta \, , \,  B \Omega_\beta) }
is analytic in the strip $- \beta / 2 < \Im z < 0$. 
Both functions are bounded and have continuous 
boundary values for $\Im z \searrow 0$ and $\Im z \nearrow 0$, respectively. Locality implies
\# { 
\lim_{\Im z \searrow 0}  (  \Omega_\beta \, , \,  
B  {\rm e}^{ iz  H_\beta} A \Omega_\beta) 
= \lim_{\Im z \nearrow 0}  ( {\rm e}^{  i \bar z H_\beta} A^* \Omega_\beta \, , \,  
B \Omega_\beta)
\qquad \forall |t| < \delta.}
Applying the Edge-of-the-Wedge Theorem [SW] we conclude that there exists a function 
\# { f_{A,B} \colon G_{\delta} \to \C ,} 
analytic on the doubly cut strip
$ G_{\delta} = \{ z \in \C : | \Im z | < \beta /2 \} 
\backslash \{ t \in \R : |t| \ge \delta \} $ such that
\# {f_{A,B} (z) = 
\left\{
\eqalign{
&  {  (  \Omega_\beta \, , \,  
B  {\rm e}^{ iz  H_\beta} A \Omega_\beta) }
\cr
&  {  ( {\rm e}^{  i \bar z H_\beta} A^* \Omega_\beta \, , \,  
B \Omega_\beta) }}
\right\} 
{\rm \ for \ }  
\left\{
\eqalign{
& { 0  < \Im z   <  \beta / 2 ,}  
\cr
& { - \beta /2  < \Im z  <  0 .}}  
\right\} 
}
The absolute value of $f_{A,B}$ at the origin can 
be estimated from the values $f_{A,B}$ takes at the boundaries:
\& {   | (\Omega_\beta \, , \,  B A \Omega_\beta) | 
& \le  \inf_{0 < \lambda < \beta /2 } 
\Bigl( \, \sup_{|t| \ge \delta} \bigl| f_{A,B} (t \pm i0) \bigr|^{1- k } 
\cdot \sup_{t \in \r} \bigl| f_{A,B} (t + i\lambda) \bigr|^{k \over 2}
\cdot \sup_{t \in \r} \bigl| f_{A,B} (t - i\lambda) \bigr|^{k \over 2} \Bigr) 
\cr
&\le  \inf_{ 0 < \lambda   < \beta / 2} 
\Bigl( \| \Omega_\beta \|^2 \cdot \|  A \| \, \|  B \| \Bigr)^{1-k}
\times
\cr
& \qquad 
\times \Bigl( \| \Omega_\beta \|^2   \cdot \|  B \|^2  \cdot
 \| {\rm e}^{-  {\lambda} H_\beta}  A \Omega_\beta \| 
\cdot \| {\rm e}^{-  {\lambda}  H_\beta} A^* \Omega_\beta \|  \Bigr)^{k / 2} }
where  $k   = {2 \over \pi} \arctan \Bigl( 2  \sinh  {   \pi \delta \over 2 \lambda} \Bigr)$.
Taking the supremum over the unit ball for $ B \in {\cal R}_\beta (\O_2)'$
and putting $\lambda = \beta / 4 $ we obtain, 
for $\|  A \| \le 1$,  
\# { \| \Xi_{\beta, *} ( A \pm  A^*) \| \le  
{\rm const} \cdot \| {\rm e}^{-  {\beta \over 4} H_\beta} ( A \pm A^*)
\Omega_\beta \|^{ {2 \over \pi} \arctan \bigl( 2  \sinh  {2 \pi \delta \over \beta} \bigr)} .}
By assumption, $\Theta_{\lambda, {\cal O}_1}$ is of order $q = 0$, thus (66) implies that
$\Xi_{*, \beta}$ is of order
$ q_* = 0 $, too [BD'AL b][BY].
Since the real linear maps $ A \mapsto ( A \pm  A^*)$ are bounded,
we conclude that $\Xi_{ \beta, *} $ is nuclear.}

\Cor{There exist non-trivial subrepresentations $\hat {\pi}$ of $\pi$ and
$\hat{\pi}_{p}$ of~$\pi_p$, respectively, which are unitarily equivalent.}

\Pr{As noted in [BD'AL a], the nuclearity of the map 
$\Xi_{ \beta, *}$ simply means that there exist 
sequences $\phi_i \in {\cal R}_\beta ( \O_1 )_*$ and  $\psi_i 
\in {\cal R}_\beta (\O_2)'_*$  with $\sum \| \phi_i \| \, \| \psi_i \| < \infty$
such that 
\# { \bigl( \Omega_\beta \, , \,  \pi (A \odot B) \Omega_\beta \bigr) 
= \sum_i \phi_i ( A) \psi_i ( B)  \qquad \forall A \in {\cal R}_\beta (\O_1), \quad  
B \in {\cal R}_\beta ( \O_2)'.}
As an absolutely convergent sum of normal functionals 
\# { \sum_i \phi_i \odot \psi_i ( \, . \,) \colon {\cal C}( \O_1 , \O_2) \to \C}
itself is, w.r.t.\ the representation $\pi_p$,
a normal\footnote{$^\dagger$}{\sevenrm A linear functional on
$\scriptstyle {\cal R}_\beta (\O_1) \odot {\cal R}_\beta (\O_2)'$ is said to 
be normal relative to $\scriptstyle \pi_p$, if it is continuous with respect to 
the ultra-weak topology determined by $\scriptstyle \pi_p$.} functional
on the algebraic tensor product~${\cal C}_\beta (\O_1, \O_2)$.
Now the algebraic tensor product is weakly dense in the $W^*$-tensor product. It follows that 
the functional 
$\bigl( \Omega_\beta \, , \pi ( \, . \,) \Omega_\beta \bigr)$ allows a unique 
continuous extension to a normal state 
on the $W^*$-tensor product ${\cal R}_\beta (\O_1) \otimes {\cal R}_\beta (\O_2)'$, 
which will be denoted 
\# { \omega_\otimes  ( \, . \, )  := \sum_i \phi_i ( \, . \, ) \otimes \psi_i ( \, . \, )  .}
Consequently, the representations $\pi$ and $\pi_p$ can not be
disjoint.} 

\Th{Let $\hat{\pi}$ and $\hat{\pi}_p$ be two
arbitrary subrepresentations of $\pi$ and $\pi_p$; respectively.
It follows that
\vskip .2cm
(i) the restrictions of $\hat {\pi}$ and $\pi$ to 
${\cal C}_\beta (\O, \hat{\O})$
are unitarily equivalent; 
\vskip .2cm
(ii) the restrictions of $\hat{\pi}_p$ and $\pi_p$ to 
${\cal C}_\beta (\O, \hat{\O})$
are unitarily equivalent.}

Combining this theorem with Corollary~3.3 we arrive at

\Th{The restrictions of $\pi$ and $\pi_p$ to 
${\cal C}_\beta (\O, \hat{\O})$
are unitarily equivalent.}

In order to prove (i) of Theorem 4.4, we need the following

\Lm{Let $E$ denote the projection onto the subspace 
${\cal K}_\beta \subset \H_\beta$ reducing~$\pi$ to $\hat{\pi}$.
It follows that $\Omega_\beta$ is cyclic for 
$\pi \bigl({\cal C}_\beta (\O, \hat{\O}) \bigr)''$ and
$E \Omega_\beta$ is separating for $\pi \bigl({\cal C}_\beta (\O, \hat{\O}) \bigr)''$.}

\Pr{Since $\O \subset \R^4$ is by assumption open and
$ {\cal R}_\beta (\O)  \subset \pi \bigl({\cal C}_\beta (\O, \hat{\O}) \bigr)'' $, 
the first part of the statement is a direct consequence of the Reeh--Schlieder property. 
The second part can be seen as follows: From the inclusions
\# {\O \subset \subset \O_1 \subset \subset \O_2  \subset \subset \hat{\O}}
it follows that
\# { \pi \bigl({\cal C}_\beta (\O, \hat{\O})\bigr)' \cap 
\pi \bigl({\cal C}_\beta (\O_1, \O_2) \bigr) \supset
{\cal R}_\beta (\O)' \cap {\cal R}_\beta (\O_1). }
By the Reeh--Schlieder theorem $\Omega_\beta$ is cyclic for 
${\cal R}_\beta (\O)' \cap {\cal R}_\beta (\O_1)$, since by assumption the closure of the open 
and bounded region $\O$ lies inside the interior of the region $\O_1$. 
Thus $\Omega_\beta$ is separating for
\# {\pi \bigl( {\cal C}_\beta (\O, \hat{\O}) \bigr)'' \vee
\pi \bigl( {\cal C}_\beta (\O_1, \O_2) \bigr)'.}
By definition, the subspace ${\cal K}_\beta := E \H_\beta$ and its orthogonal
complement ${\cal K}_\beta^\bot$ are invariant under the action of
$\pi \bigl({\cal C}_\beta (\O_1, \O_2) \bigr)$. By standard arguments it follows that
$E \in \pi \bigl( {\cal C}_\beta (\O_1, \O_2) \bigr)'$.
Hence, if $Z E \Omega_\beta = 0$ for some projection\footnote{$^\dagger$}{\sevenrm
Given an arbitray 
element $\scriptstyle C \in \pi ({\cal C}_\beta (\O, \hat{\O}) )'' $ 
one can use the spectral 
decomposition of $\scriptstyle C^*C$ in order to reduce the general case 
to the case of projections:
With $\scriptstyle C^*C$ also the spectral
projections of $\scriptstyle C^*C$ belong to 
$\scriptstyle \pi ({\cal C}_\beta (\O, \hat{\O}) )''$; 
and obviously $\scriptstyle C^*C = 0$ implies $\scriptstyle C=0$. }
$ Z \in 
\pi \bigl({\cal C}_\beta (\O, \hat{\O}) \bigr)'' $,  
then (72) implies $Z E  = 0$. Because of locality 
\# { [ {\rm e}^{iH_\beta t} Z {\rm e}^{-iH_\beta t} , E ] = 0  \qquad \forall t \in {\cal U},}
where ${\cal U}$ denotes some open neighborhood of the origin in $\R$.
Since $\pi \bigl({\cal C}_\beta (\O_1, \O_2) \bigr)' \subset {\cal R}_\beta$
the thermal version [J\"a a] of a classical Lemma by Borchers [Bo] applies 
and yields 
\# { E {\rm e}^{iH_\beta t}  Z = 0 \qquad \forall t \in \R.}
By assumption, $\Omega_\beta$ is the unique --- up to a phase ---
normalized eigenvector for the only discrete eigenvalue $\{ 0 \}$ of $H_\beta$,
thus 
\& { 0 & = \lim_{T \to \infty} {1 \over 2T} \int_{-T}^{T} {\rm d} t \, 
(\Omega_\beta \, , \, E {\rm e}^{it H_\beta} Z \Omega_\beta )
\cr
& =  
(\Omega_\beta \, , \, E \Omega_\beta ) ( \Omega_\beta \, , Z \Omega_\beta ) 
= \| E \Omega_\beta \|^2 \, \| Z \Omega_\beta \|^2.}
By definition, $E \Omega_\beta \ne 0$, thus (75) implies 
$Z \Omega_\beta = 0$. $\Omega_\beta$ is separating for
$\pi \bigl({\cal C}_\beta (\O, \hat{\O}) \bigr)'' $, thus $Z = 0$. 
This proves that the vector $E \Omega_\beta$ is separating 
for~$\pi \bigl( {\cal C}_\beta (\O, \hat{\O} ) \bigr)''$.}

\Cor{Let $E$ denote the projection onto the subspace 
${\cal K}_\beta \subset \H_\beta$ reducing~$\pi$ to~$\hat{\pi}$. It follows that
$E \in \pi \bigl( {\cal C}_\beta (\O_1, \O_2) \bigr)'$ can be represented in the form 
\# { E = V V^*, \qquad \hbox{where} \qquad V \in
\pi \bigl( {\cal C}_\beta (\O , \hat{\O}) \bigr)' }
is an isometry, i.e., $V^* V = \1 $.}

\Pr{$\Omega_\beta$ is cyclic for 
$\pi \bigl( {\cal C}_\beta (\O , \hat{\O}) \bigr)''$ 
and in addition has the property that $E \Omega_\beta$ is separating for 
$\pi \bigl( {\cal C}_\beta (\O , \hat{\O}) \bigr)''$. It follows that
$(E \Omega_\beta, \, . \, E \Omega_\beta)$ defines a faithful normal state
on $\pi \bigl( {\cal C}_\beta (\O , \hat{\O}) \bigr)''$. Moreover,
$\pi \bigl( {\cal C}_\beta (\O , \hat{\O}) \bigr)''$ has a 
cyclic and separating vector, namely~$\Omega_\beta$.
We conclude (see e.g.\ [Sa, 2.7.9] or [BR, 2.5.31])
that there exists another vector $\Psi \in \H_\beta$, cyclic and separating for 
$\pi \bigl( {\cal C}_\beta (\O , \hat{\O}) \bigr)''$, which satisfies
\# { (\Psi \, , \pi (C^*C) \Psi) = (E \Omega_\beta, \pi (C^*C) E \Omega_\beta) \qquad \forall C
\in {\cal C}_\beta (\O , \hat{\O}).}
Taking into account  the properties of $\Omega_\beta$ and $\Psi$ and
\# { E \in \pi \bigl( {\cal C}_\beta (\O_1, \O_2) \bigr)'
\subset \pi \bigl( {\cal C}_\beta (\O , \hat{\O}) \bigr)' }
it follows that  
\# { V \pi (C) \Psi = \pi (C) E \Omega_\beta \qquad \hbox{for} \qquad C
\in {\cal C}_\beta (\O , \hat{\O}) }
defines an isometry $V$ with the desired properties. }

\Rem{The isometry $V \colon \H_\beta \to {\cal K}_\beta$ satisfies
$V^* V = {\1 }_{\H_\beta}$ and $V V^* = {\1 }_{{\cal K}_\beta}$. 
It therefore establishes the unitary equivalence between
the restrictions of $\pi$ and $\hat{\pi}$ to 
${\cal C}_\beta (\O, \hat{\O}) $.}

The proof of part (ii) of Theorem 3.4 follows the same line of arguments:
We  show that $\Omega_\beta \otimes \Omega_\beta \in \H_\beta \otimes 
\H_\beta$ is cyclic for 
$\pi_p \bigl( {\cal C}_\beta (\O, \hat{\O}) \bigr)''$ 
and in addition has the property that $E_{p} (\Omega_\beta \otimes \Omega_\beta)$ 
is separating for $\pi_p \bigl( {\cal C}_\beta (\O , \hat{\O}) \bigr)''$, where
$E_{p}$ denotes the projection onto the subspace 
${\cal K}_p \subset \H_\beta \otimes \H_\beta$
reducing~$\pi_p$ to $\hat{\pi}_p$. In order to do so, we
adapt the classical lemma of Borchers cited above to the tensor product 
representation.

\Lm{Let $P \in {\cal R}_\beta \otimes \B(\H_\beta)$ and let
$Q \in \B(\H_\beta) \otimes {\cal R}_\beta$ be a (self-adjoint) projection operator such that
\# { Q P  = 0 \qquad \hbox{and} \qquad [U_p (t) Q  U_p (-t) \, , \, P] = 0
\qquad \forall |t| < \delta ,}
where $U_p \colon \R \to \B(\H_\beta) \otimes \B(\H_\beta)$ is given by
$t \mapsto {\rm e}^{ i t H_\beta} \otimes {\rm e}^{ i t H_\beta} $ and $\delta > 0$. 
It follows that
\# { \Bigl( \Omega_\beta \otimes \Omega_\beta \, , \, Q U_p (t) P 
(\Omega_\beta \otimes \Omega_\beta) \Bigl) = 0 \qquad \forall t \in \R .}
}

\Pr{Due to the KMS relation, the function
\# { f^+ (z) :=  
\Bigl(  \bigl( \1 \, \otimes {\rm e}^{ - i \bar z  H_\beta} \bigr) Q^* \bigl( \Omega_\beta  
\otimes \Omega_\beta \bigr) \, , \,  
\bigl( {\rm e}^{ iz  H_\beta} \otimes \1 \bigr) P \bigl(\Omega_\beta 
\otimes \Omega_\beta \bigr) \Bigr) }
is analytic in the strip $S(0, \beta/2) := \{ z \in \C : 0 < \Im z < \beta/2 \}$, while the function
\# { f^- (z) :=
\Bigl( \bigl( {\rm e}^{ i \bar z  H_\beta} \otimes \1 \bigr) \, P^* \bigl(\Omega_\beta 
\otimes \Omega_\beta \bigr) \, , \,  
\bigl( \1 \, \otimes {\rm e}^{ - i z  H_\beta} \bigr) Q  \bigl( \Omega_\beta \otimes 
\Omega_\beta \bigr) \Bigr) }
is analytic in the strip $S(-\beta/2, 0) := \{ z \in \C : - \beta / 2 < \Im z < 0 \}$. 
Both functions are bounded and have continuous 
boundary values for $\Im z \searrow 0$ and $\Im z \nearrow 0$, respectively.
Now (80) implies
\& { 
\lim_{\Im z \searrow 0}  & \Bigl( \bigl(  \1 \otimes {\rm e}^{-i \bar z  H_\beta} \bigr)
Q^* \bigl( \Omega_\beta \otimes \Omega_\beta \bigr) \, , \,  
\bigl( {\rm e}^{ i z  H_\beta} \otimes \1 \bigr) P \bigl( \Omega_\beta \otimes \Omega_\beta 
\bigr) \Bigr)
\cr
&=    \Bigl( 
\Omega_\beta \otimes \Omega_\beta   \, , \,  Q U_p ( \Re z) P 
\bigl( \Omega_\beta \otimes \Omega_\beta \bigr) \Bigr)
=    \Bigl( 
\Omega_\beta \otimes \Omega_\beta   \, , \,  P U_p (- \Re z) Q 
\bigl( \Omega_\beta \otimes \Omega_\beta \bigr) \Bigr)
\cr
&= \lim_{\Im z \nearrow 0}   \Bigl( \bigl( {\rm e}^{ i \bar z  H_\beta} \otimes \1 \bigr)
P^* \bigl( \Omega_\beta \otimes \Omega_\beta \bigr) \, , \,  
\bigl( \1 \otimes {\rm e}^{ - i z  H_\beta} \bigr) Q^* 
\bigl( \Omega_\beta \otimes \Omega_\beta \bigr) \Bigr)
\quad  \forall |\Re z| < \delta.}
Using the Edge-of-the-Wedge Theorem  one concludes that there exists a function 
\# { f_{P, Q} \colon G_{\delta} \to \C } 
which is analytic on the doubly cut strip
\# { G_{\delta} = \{ z \in \C : - \beta / 2  < \Im z  < \beta / 2 \} \setminus \{ z \in \C : 
\Im z = 0, |\Re z| \ge \delta \}   }
and satisfies
\# {f_{P, Q} (z) = 
\left\{
\eqalign{
&  {  f^+ (z)  } \cr
&  {  f^- (z)  }}
\right\} 
{\rm \ for \ }  
\left\{
\eqalign{
& { 0  < \Im z   <  \beta / 2 ,}  
\cr
& { - \beta /2  < \Im z  <  0 .}}  
\right\} 
}
By assumption $Q P = 0$, hence $f_{P,Q} (0) = 0$.
According to Lagrange's theorem $f_{P,Q}$ vanishes identically, 
if $0$ is a zero of infinite order. This follows from the original 
arguments of Borchers: set
\# { t^{(i)}_j := {\delta j \over 2 i n} , \qquad i \in \N, \quad j = \{ 1, \ldots , n \} ,}
and $Q \bigl( t^{(i)}_j \bigr) := U_p \bigl( t^{(i)}_j \bigr) Q  U_p \bigl(-t^{(i)}_j \bigr)$.
It follows that
\# { \bigl[  P \, , \,  U_p(t)  Q \bigl( t^{(i)}_1 \bigr)     
\ldots Q \bigl( t^{(i)}_n \bigr) U_p(-t) \bigr] = 0 \qquad 
\forall |t| < \delta / 2.}
The functions $f_{t^{(i)}_1, \ldots, t^{(i)}_n}^+ \colon S(0, \beta/2) \to \C$,
\# {
z \mapsto  \Bigl( 
\bigl(  \1 \otimes {\rm e}^{-i \bar z  H_\beta} \bigr)
Q^* \bigl(t^{(i)}_n \bigr)  \ldots Q^* \bigl( t^{(i)}_1 \bigr) \bigl( \Omega_\beta \otimes 
\Omega_\beta \bigr) \, , \, \bigl( {\rm e}^{iz H_\beta} \otimes \1 \bigr) P 
\bigl( \Omega_\beta \otimes 
\Omega_\beta \bigr) \Bigr)}
and $f_{t^{(i)}_1, \ldots, t^{(i)}_n}^- \colon S (- \beta /2 , 0) \to \C$,
\# { z \mapsto
\Bigl( \bigl( {\rm e}^{i \bar z H_\beta} \otimes \1 \bigr) P^* 
\bigl( \Omega_\beta \otimes \Omega_\beta \bigr) \, , \, \ 
\bigl( \1 \otimes {\rm e}^{- i z H_\beta} \bigr)
Q \bigl( t^{(i)}_1 \bigr) \ldots Q \bigl( t^{(i)}_n \bigr)  \bigl( \Omega_\beta \otimes 
\Omega_\beta \bigr) \Bigr)  } 
are bounded, analytic in the interior of their domains and continuous at the 
boundary. The boundary values for $ \Im z \searrow 0$ resp.\ 
$\Im z \nearrow 0 $ coincide for $ | \Re z | < 
\delta / 2 $. 
Applying the Edge-of-the-Wedge Theorem [SW] one concludes
that the functions defined in (90) and (91) are  
the restrictions to the upper (resp.\ lower)
half of the doubly cut strip ${\cal G}_{\delta /2}$ 
of a function 
\# {
f_{t^{(i)}_1, \ldots, t^{(i)}_n} (z) 
:= 
\left\{
\eqalign{
&  { f_{t^{(i)}_1, \ldots, t^{(i)}_n}^+ (z) }
\cr
&
{ f_{t^{(i)}_1, \ldots, t^{(i)}_n}^- (z) } 
}
\right\} 
\hbox{\rm for}
\left\{
\eqalign{
& { 0  < \Im z < \beta/2,}  
\cr
& {-\beta/2 < \Im z < 0 ,}  }  
\right\} 
}
defined and analytic for $z \in {\cal G}_{\delta /2}$.
The function 
$ f_{t^{(i)}_1, \ldots, t^{(i)}_n} $ has continuous boundary values for $z \to \partial 
{\cal G}_{\delta /2}$, uniformly bounded by one: 
For example,
\& {
\sup_{ s \in \r }  f_{t^{(i)}_1,  \ldots, t^{(i)}_n} (s + i \beta/2)   
& \le \Bigl\| \bigl( \1 \otimes J {\rm e}^{ -{\beta \over 2} H_\beta} \bigr)
Q^* \bigl(t^{(i)}_1\bigr) \ldots Q^* \bigl(t^{(i)}_n \bigr)
\bigl( \Omega_\beta \otimes \Omega_\beta \bigr)  \Bigr\|
\times
\cr
&
\qquad 
\qquad 
\qquad 
\times
 \Bigl\| \bigl( J {\rm e}^{- {\beta \over 2} H_\beta} \otimes \1 \bigr)
 P   \bigl( \Omega_\beta \otimes \Omega_\beta \bigr) \Bigr\| .}
Note that $P=P^*$ implies
\& { \bigl( J {\rm e}^{- {\beta \over 2} H_\beta} \otimes \1 \bigr)
 P   \bigl( \Omega_\beta \otimes \Omega_\beta \bigr)
&
= \sum_i J {\rm e}^{- {\beta \over 2} H_\beta} P^{(1)}_i \Omega_\beta
\otimes P^{(2)}_i \Omega_\beta
\cr 
&
= \sum_i \bigl(P^{(1)}_i\bigr)^* \Omega_\beta
\otimes P^{(2)}_i \Omega_\beta
= P \bigl( \Omega_\beta \otimes \Omega_\beta \bigr).}
The same argument applies to the first term on the r.h.s.\ in (93).  
Moreover, $\| \Omega_\beta \| = 1$, $\| P \| = 1$, $\| Q \| = 1$, and
$\| U_p (s) \| = 1$ for all $s \in \R$.
By an application of the maximum modulus 
principle we obtain  
\# {  \bigl| f_{t^{(i)}_1, \ldots, t^{(i)}_n} (z)  \bigr|  \le 1 \qquad
\forall z \in {\cal G}_{\delta / 2}.}
By assumption $Q P = 0$, hence
\# { f_{t^{(i)}_1, \ldots, t^{(i)}_n} \bigl( - t^{(i)}_j \bigr) = 0 .}
We conclude that inside the circle  $|z| < \delta / 2$ each of the
functions  $f_{t^{(i)}_1, \ldots, t^{(i)}_n}$ possesses $n$ zeros for
pairwise different values of $t^{(i)}_j$. Thus all of the functions
\# { g_{t^{(i)}_1, \ldots, t^{(i)}_n} (z) := 
{f_{t^{(i)}_1, \ldots, t^{(i)}_n} (z)  \over \prod_{j = 1}^n \bigl( z + t^{(i)}_j \bigr)} ,
\qquad i \in \N, \quad j = \{ 1, \ldots, n \}, }
are analytic in the open disc $D_{\delta / 2}$ of radius $\delta / 2 $ and
centered at the origin. Note that by definition 
$D_{ \delta / 2 } \subset {\cal G}_{ \delta / 2 }$.
Yet the number of zeros does not change in the limit $t^{(i)}_j \to 0$ and consequently,
for $i >1$,
\# { \Bigl| g_{t^{(i)}_1, \ldots, t^{(i)}_n} (z) \Bigr| 
\le
\sup_{w \in \partial D_{ \delta / 2 }} 
{  | f_{t^{(i)}_1, \ldots, t^{(i)}_n} (w) | \over  \prod_{j = 1}^n \bigl| w + t^{(i)}_j
\bigr|  } 
\le \Bigl({ 4  \over  \delta }\Bigr)^n  
\qquad \forall z \in D_{ \delta / 2 }.}
In the last inequality we used $\bigl| w + t_j^{(i)} \bigr| \ge \bigl| |w| - |t_j^{(i)}| \bigr|$ and
$|w| = \delta / 2 $ together with $\bigl| t_j^{(i)} \bigr| < \delta / 4$ for $i > 1$.
Hence,
\# {   \bigl| f_{ t^{(i)}_1, \ldots, t^{(i)}_n } (z)  \bigr|       
\le \Bigl({ 4  \over  \delta }\Bigr)^n  
\prod_{j = 1}^n \bigl| z + t^{(i)}_j \bigr|
\le  {\rm const} \cdot  |z|^n 
\qquad \forall z \in D_{ \delta / 2 }.}
Because of $Q^2 = Q$,  
$ f_{0, \ldots, 0} $ coincides with $f_{P, Q} $.  
The group $t \mapsto U_p (t)$ is  strongly continuous, thus 
\# { \bigl| f_{t^{(i)}_1, \ldots, t^{(i)}_n} (z) - f_{0, \ldots, 0} (z)  \bigr| \to 0 
\qquad {\rm for} \quad
t^{(i)}_j \to 0 , \quad j= 1, \ldots ,n,  }
uniformly in $z \in G_{ \delta / 2 }$. Hence $0$ is a zero of $n$-th order:
\# { | f_{P, Q} (z) | \le  {\rm const} \cdot |z|^n  
\qquad \forall z \in D_{ \delta / 2 } .}
Since $n \in \N$ was arbitrary, we conclude that
$f_{P, Q}$ vanishes identically. }

\Lm{The vector $\Omega_\beta \otimes \Omega_\beta \in \H_\beta \otimes \H_\beta$ is cyclic for 
$\pi_p \bigl({\cal C}_\beta (\O, \hat{\O}) \bigr)''$ and
$E_p (\Omega_\beta \otimes \Omega_\beta)$ is separating 
for~$\pi_p \bigl({\cal C}_\beta (\O, \hat{\O}) \bigr)''$, 
where $E_p \in \pi_p \bigl({\cal C}_\beta (\O_1, \O_2) \bigr)'$ 
denotes the projection onto the subspace 
${\cal K}_p \subset \H_\beta \otimes \H_\beta$ reducing~$\pi_p$.}

\Pr{By the Reeh--Schlieder theorem $\Omega_\beta
\otimes \Omega_\beta$ is cyclic for 
\# { {\cal R}_\beta (\O)' \otimes {\cal R}_\beta (\hat{\O})
\subset \pi_p \bigl({\cal C}_\beta (\O, \hat{\O}) \bigr)'' .} 
By assumption
\# { \O \subset \subset \O_1 \subset \subset \O_2 \subset \subset \hat{\O} .}
It follows from the general theory of intersections of $W^*$-tensor products [Ta] that
\& { \pi_p \bigl( {\cal C}_\beta ( \O, \hat{\O} ) \bigr)' \cap 
\pi_p \bigl( {\cal C}_\beta (\O_1, & \O_2) \bigr)'' \supset
\Bigl({\cal R}_\beta (\O) \otimes {\cal R}_\beta (\hat{\O})' \Bigr)' 
\cap  \Bigl( {\cal R}_\beta (\O_1) \otimes {\cal R}_\beta (\O_2)' \Bigr)  
\cr
& = \Bigl( {\cal R}_\beta (\O)' \cap {\cal R}_\beta (\O_1) \Bigr)
\otimes 
\Bigl( {\cal R}_\beta (\hat{\O}) \cap {\cal R}_\beta (\O_2)' \Bigr)
\cr
& \supset \Bigl( {\cal R}_\beta (\O') \cap {\cal R}_\beta (\O_1) \Bigr)
\otimes 
\Bigl( {\cal R}_\beta (\hat{\O}) \cap {\cal R}_\beta (\O_2') \Bigr). }
By assumption, both 
\# {\O' \cap \O_1 \qquad \hbox{\rm and} \qquad \hat{\O} \cap \O_2' }
contain open subsets. Thus, due to 
the Reeh--Schlieder property, $\Omega_\beta \otimes \Omega_\beta$ is cyclic for
$\pi_p \bigl( {\cal C}_\beta ( \O, \hat{\O} ) \bigr)' \cap 
\pi_p \bigl( {\cal C}_\beta (\O_1, \O_2) \bigr)''$ and therefore separating for
\# { \pi_p \bigl( {\cal C}_\beta (\O, \hat{\O}) \bigr)'' \vee
\pi_p \bigl( {\cal C}_\beta (\O_1, \O_2) \bigr)'. }
Now let $ Z_p \in 
\pi_p \bigl({\cal C}_\beta (\O, \hat{\O}) \bigr)'' =  {\cal R}_\beta (\O) 
\otimes {\cal R}_\beta (\hat{\O})' $ be some projection such that 
\# { Z_p E_p \bigl( \Omega_\beta \otimes \Omega_\beta \bigr) = 0. }
Since $E_p \in \pi_p \bigl( {\cal C}_\beta (\O_1, \O_2) \bigr)'$,
it follows that $Z_p E_p \in \pi_p \bigl({\cal C}_\beta (\O, \hat{\O}) \bigr)'' \vee
\pi_p \bigl( {\cal C}_\beta (\O_1, \O_2) \bigr)'$ and consequently
$Z_p E_p = 0$. Moreover,
\# { \bigl[ U_p (t) Z_p U_p (-t) , E_p \bigr] = 0  \qquad \forall t \in {\cal U},}
where ${\cal U}$ denotes some open neighborhood of the origin in $\R$. According to
Lemma 4.8 $Z_p E_p  = 0$ now implies   
\# { \Bigl( \Omega_\beta \otimes \Omega_\beta \, , \, Z_p U_p (t) E_p 
(\Omega_\beta \otimes \Omega_\beta) \Bigl) = 0 \qquad \forall t \in \R .}
Now $\Omega_\beta \otimes \Omega_\beta$ is the unique --- up to a phase ---
normalized, invariant eigenvector for the one-parameter group $t \mapsto U_p (t) $.
Thus, by the same argument as in the proof of Lemma~3.6,
\# { E_p \bigl( \Omega_\beta \otimes \Omega_\beta \bigr) \ne 0 
\Rightarrow Z_p \bigl( \Omega_\beta \otimes \Omega_\beta \bigr) = 0,}
which is only possible if $Z_p = 0$. It follows that
the vector $E_p \bigl( \Omega_\beta \otimes \Omega_\beta \bigr)$ 
is separating for~$\pi_p \bigl( {\cal C}_\beta (\O, \hat{\O} ) \bigr)''$.}

\Cor{Let $E_p$ denote the projection onto the subspace 
${\cal K}_p \subset \H_\beta \otimes \H_\beta$ reducing~$\pi_p$ to~$\hat{\pi}_p$. It follows that
$E_p \in \pi_p \bigl( {\cal C}_\beta (\O_1, \O_2) \bigr)'$ can be represented in the form 
\# { E_p = V_p V_p^*, \qquad \hbox{where} \qquad V_p \in
\pi_p \bigl( {\cal C}_\beta (\O , \hat{\O}) \bigr)' }
is an isometry, i.e., $V_p^* V_p = { \1 }_{\H_\beta \otimes \H_\beta}$ and 
$V_pV_p^* = { \1 }_{{\cal K}_p}$.}

The proof of this result is --- up to notation --- identical with the proof of
Corollary~3.7, therefore we do not repeat the argument.

\vskip .5cm

We summarize our result in the following

\Th{Assume a TFT is specified by a net
\# { \O \to {\cal R}_\beta (\O), \qquad \O \subset \R^4,}
of von Neumann algebras, subject to the standard assumptions stated explicitly
on p.6. Furthermore assume that for any bounded space--time region $\O$ 
the maps $\Theta_{\lambda, {\cal O}} \colon {\cal R}_\beta (\O) \to  \H_\beta$,
\# {  A  \mapsto {\rm e}^{- \lambda | H_\beta |} A \Omega_\beta , }
are of type $s$ (order 0) for any $\lambda > 0$. 
It follows that for any inclusion of open bounded space--time regions 
$\O \subset \subset \hat{\O}$, there exists a type I 
factor~${\cal N}_\beta (\O, \hat{\O})$ such that
\# { {\cal R}_\beta (\O)  \subset {\cal N}_\beta (\O, \hat{\O})
\subset {\cal R}_\beta (\hat{\O})  ,}
provided the closure of $\O$ is contained in the interior of $\hat {\cal O}$.}

\Pr{Theorem 4.5 ensures that there exists a unitary operator $W$ 
mapping $\H_\beta$ onto $ \H_\beta \otimes \H_\beta$ such that
\# { WABW^{-1} = A \otimes B }
for all $A \in {\cal R}_\beta (\O_1)$ and all $B \in {\cal R}_\beta (\O_2)'$.
 Set  ${\cal N}_\beta := W^* \bigl( \B(\H_\beta) \otimes \1 \, \bigr) W$. 
Clearly ${\cal N}_\beta$ is 
a type I factor and since there holds the trivial inclusion
\# { W^{-1} \Bigl( {\cal R}_\beta (\O) \otimes \1 \, \Bigr) W 
\subset  W^{-1} \Bigl( \B(\H_\beta) \otimes \1 \Bigr) \, W \subset 
\Bigl( W^{-1}  \bigl( \1 \otimes {\cal R}_\beta (\O_2) \bigr) W \Bigr)', } 
we arrive at (114).}

\vskip 1cm
\Hl{Equivalent Formulations of the Split Property}

\noindent
We start with the following  

\Th{Let $\O$  be a bounded space--time region such that 
the closure of $\O$ is contained in the interior of $\hat {\cal O}$.
Then the following five conditions are equivalent:
\vskip .3cm
\halign{ \indent #  \hfil & \vtop { \parindent = 0pt \hsize=34em
                            \strut # \strut} 
\cr 
(i)    & (Split property).
There exists a type I factor ${\cal N}_\beta$ such that 
\# {{\cal R}_\beta (\O) \subset {\cal N}_\beta \subset{\cal R}_\beta (\hat {\cal O}) .}
\cr
(ii)    & (Existence of normal product state extensions for partial states).
For any pair of normal states $\phi_1$ of ${\cal R}_\beta (\O)$ and
$\phi_2$ of ${\cal R}_\beta (\hat {\cal O})'$ there exists a normal
state $\phi_{1,2}$ on $\B(\H_\beta)$ which is an extension of 
$\phi_1$ and $\phi_2$ and a product state for the 
von Neumann algebras ${\cal R}_\beta (\O)$ 
and ${\cal R}_\beta (\hat {\cal O})'$.
\cr
(iii)    & (Existence of a normal product state).
There exists a normal state~$\phi$ on $\B(\H_\beta)$ which is a product state for the 
von Neumann algebras ${\cal R}_\beta (\O)$ and ${\cal R}_\beta (\hat {\cal O})'$.
\cr
(iv)    & (Existence of a faithful normal product state
extension of the KMS state).
There exists a normal product state~$\omega_p$ on ${\cal R}_\beta (\O) 
\vee {\cal R}_\beta (\hat {\cal O})'$ such that
\# {\omega_p ( A B ) = (\Omega_\beta , A \Omega_\beta)  (\Omega_\beta, B \Omega_\beta)}
for all $A \in {\cal R}_\beta (\O)$ and
$B \in {\cal R}_\beta (\hat {\cal O})'$.  Moreover, 
$\omega_p$ is faithful on the von Neumann algebra 
${\cal R}_\beta (\O) \vee {\cal R}_\beta (\hat {\cal O})'$.
\cr
(v)   & (Canonical cyclic and separating product vector). 
There exists a unique vector $\eta \in \H_\beta$ in the natural positive cone
${\cal P}^\natural_{\Omega_\beta} 
\bigl({\cal R}_\beta (\O) \vee {\cal R}_\beta (\hat {\cal O})'\bigr)$ such that
\vskip .1cm
\hskip .5cm
a.) $(\eta, A B \eta) = (\Omega_\beta , A \Omega_\beta)  (\Omega_\beta, B \Omega_\beta)$
for all $A \in {\cal R}_\beta (\O)$ and
$B \in {\cal R}_\beta (\hat {\cal O})'$.
\vskip .1cm
\hskip .5cm
b.) $\eta$ is cyclic and separating for ${\cal R}_\beta (\O) \vee {\cal R}_\beta ( \hat {\cal O})'$.
\cr
(vi)    & (Statistical independence). The von Neumann algebra generated by ${\cal R}_\beta (\O) $
and ${\cal R}_\beta (\hat {\cal O})' $ is isomorphic to the 
$W^*$-tensor product of the two algebras.
This means that there exists a unitary operator 
$W \colon \H_\beta \to \H_\beta \otimes \H_\beta$  such that
\# { W A B W^* = A \otimes B } 
for all $A \in {\cal R}_\beta (\O)$ and
$B \in {\cal R}_\beta (\hat {\cal O})'$ and,  
hence, locality is reflected in an especially simple 
algebraic structure of the net $\O \to {\cal R}_\beta (\O)$.
\cr}  
}

% For convenience we reproduce (see [BR, 2.5.31]) the following 

% \Lm{Let ${\cal M}$ be a von Neumann algebra with 
% cyclic and separating vector~$\Omega$. Given a normal state
% $\phi$ on ${\cal M}$ there exists a unique vector $\chi $ 
% in the natural positive 
% cone ${\cal P}^\natural_\Omega ({\cal M}) $
% such that 
%
% \# { \phi (M) = (\chi \, , \, M \chi) \qquad \forall M \in {\cal M}.  }
%
% Moreover, if $\phi$ is faithful on ${\cal M}$, then $\chi$ is cyclic and separating for
% ${\cal M}$.} 

\Pr{i) $\Rightarrow$ ii) The KMS vector $\Omega_\beta$ is cyclic and separating for 
${\cal R}_\beta (\O) $, ${\cal R}_\beta (\hat {\cal O})$ 
and
${\cal R}_\beta (\O)' \cap {\cal R}_\beta (\hat {\cal O})$; and
therefore this $W^*$-Split-inclusion is standard. Consequently, 
the underlying Hilbert space $\H_\beta$ is separable and infinite dimensional 
[DL, Prop.\ 1.6]:
The KMS state is faithful w.r.t.\ ${\cal R}_\beta$, ${\cal N}_\beta$ is countably decomposable,
hence separable in the ultraweak topology (being of type I). It follows that 
\# { \H_\beta = \overline{ {\cal N}_\beta \Omega_\beta} }
is separable.
All infinite type I factors with infinite commutant on 
a separable Hilbert space are unitarily
equivalent to $\B(\H_\beta) \otimes \1  \, $ [KR, Ch.\ 9.3]. 
It follows that
there exists a unitary operator $W \colon \H_\beta \to \H_\beta \otimes \H_\beta$ such that 
\# { {\cal N}_\beta = W^* \bigl( \B(\H_\beta)  \otimes \1  \, \bigr) W .}
The split property (117) implies 
\# {  
W {\cal R}_\beta ( \O ) W^* \subset \B(\H_\beta) \otimes \1 \subset  
W {\cal R}_\beta (\hat {\O}) W^* , } 
and ${\cal R}_\beta ( \O )' \supset {\cal N}_\beta ' \supset  
{\cal R}_\beta (\hat {\O})'$. It follows that
\# {  
W {\cal R}_\beta ( \O )' W^* \supset \1 \otimes \B(\H_\beta) \supset  
W  {\cal R}_\beta (\hat {\O})'  W^* . } 
Let $\phi_1$ and $\phi_2$ denote two normal states over 
${\cal R}_\beta (\O)$ and ${\cal R}_\beta (\hat {\O})'$, 
respectively. Set
\# { \phi_{1,2}  := (\phi_1 \otimes \phi_2)^W, }
where
$ \phi^W (C) := \phi(W C W^*)$ for all 
$C \in {\cal R}_\beta (\O) \vee {\cal R}_\beta ( \hat {\cal O})'$. 
The state $\phi_{1,2}$ is normal and
satisfies 
\# { \phi_{1,2} (A B) = \phi_1 (A)  \phi_2 (B) }
for all $A \in {\cal R}_\beta (\O)$ and $B \in {\cal R}_\beta (\hat {\O})'$.
\vskip .2cm
\noindent
(ii) $\Rightarrow$ (iii) is trivial.
\vskip .2cm
\noindent
(iii) $\Rightarrow$ (iv) 
Let $\O_\circ$, $\O_1$, $\O_2$ and $\O_3$ denote space--time regions such that
\# {\O + te \subset \O_\circ \subset \subset \O_1 \subset \subset
\O_2 \subset \subset \O_3 + te\subset  \hat{\O}  \quad \hbox{for}
\quad |t| < \delta /3.}
From (iii) we conclude that there exists a normal product state 
$\hat \phi$ for the pair ${\cal R}_\beta (\O_1)$
and~${\cal R}_\beta (\O_2)'$. Moreover, ${\cal R}_\beta (\O_1) \vee {\cal R}_\beta (\O_2)'$
has a cyclic and separating vector, namely $\Omega_\beta$. It follows (see [BR, 2.5.31])
that there exists a vector $\hat {\xi} \in \H_\beta$ such that
\# {\hat {\phi} (C) = (\hat {\xi} \, , \, C \hat {\xi}) \qquad \forall 
C \in {\cal R}_\beta (\O_1) \vee {\cal R}_\beta (\O_2)'.}
The following argument is due to Buchholz [Bu]:
Let $P_1$, $P_2$ be the projections onto the closed subspaces
$\overline{ {\cal R}_\beta (\O_1) \hat {\xi}}$ and 
$\overline{ {\cal R}_\beta (\O_2)' \hat {\xi}}$ 
of $\H_\beta$. It is obvious that 
$P_1 \in  {\cal R}_\beta (\O_1)'$ and 
$P_2 \in {\cal R}_\beta (\O_2)$. From the 
factorization property of $\hat {\xi}$ it follows that
\# {P_1 B P_1 = (\hat {\xi} \, , B \hat {\xi}) \cdot P_1 \qquad \forall 
B \in {\cal R}_\beta (\O_2)' }
and
\# {P_2 A P_2 = (\hat {\xi} \, , A \hat {\xi}) \cdot P_2 \qquad \forall 
A \in {\cal R}_\beta (\O_1) .}
Therefore the state 
\# { \omega_1 (.) := { (P_1 \Omega_\beta \, , \, . \, P_1 \Omega_\beta)
\over \| P_1 \Omega_\beta \|^2} }
is again a product state for ${\cal R}_\beta (\O_1)$ 
and ${\cal R}_\beta (\O_2)'$. Now assume
\# { \omega_1 (A^*A) = 0 \qquad \hbox{for} \quad A \in {\cal R}_\beta (\O_\circ). }
The KMS vector $\Omega_\beta$ is separating for ${\cal R}_\beta (\O_\circ)
\vee {\cal R}_\beta (\O_1)'$, thus 
\# { A P_1 \Omega_\beta = 0  \Rightarrow A P_1 = 0.}
The Schlieder property for ${\cal R}_\beta (\O_\circ)$ and 
${\cal R}_\beta (\O_1)'$ implies $A=0$ or $P_1 = 0$. 
We conclude that $\omega_1$ is faithful for ${\cal R}_\beta (\O_\circ)$.
It follows (see [BR, 2.5.31]) that there exists a vector $\xi_1 \in \H_\beta$,
cyclic and separating  for ${\cal R}_\beta (\O_\circ)$,
which represents the restriction 
of $\omega_1$ to~${\cal R}_\beta (\O_\circ)$. Consequently,
we can construct in a canonical way an isometric operator 
$U_1 \in {\cal R}_\beta (\O_1)'$:
\# {U_1 A \xi_1 := A \cdot { P_1 \Omega_\beta \over \| P_1 \Omega_\beta \| } \qquad
\hbox {for} \quad  A \in {\cal R}_\beta (\O_\circ).}
It is evident that the range of $U_1$ equals $P_1 \H_\beta$; thus 
\# {U_1 U_1^* = P_1 \qquad {\rm and} \qquad U_1^* U_1 = \1 .}
From (134) 
and the relation $P_1 B P_1 = (\hat {\xi} \, , B \hat {\xi}) \cdot P_1 $ we get 
\& { (A \xi_1 \, , \, U_1^* B U_1 A \xi_1 ) 
& = (U_1 A \xi_1 \, , \, P_1 B P_1 U_1 A \xi_1 )
\cr & = (\hat{\xi} \, , \, B \hat{\xi}) (\xi_1 \, , \, A^* A \xi_1 ) }
The cyclicity of $\xi_1 $ w.r.t.\ ${\cal R}_\beta (\O_\circ)$ implies
\# { U_1^* B U_1 = (\hat{\xi} \, , \, B \hat{\xi}) \cdot \1 \qquad
\hbox {for} \quad  B \in {\cal R}_\beta (\O_2)'.}
Therefore the state 
\# { \omega_1 (.) := (U_1 \Omega_\beta \, , \, . \, U_1 \Omega_\beta) }
is a product state for ${\cal R}_\beta (\O_\circ)$ and 
${\cal R}_\beta (\O_2)'$ 
and the restriction of $\omega_1$ to ${\cal R}_\beta (\O_\circ)$ 
coincides with the restriction 
of the KMS state $\omega_\beta$ to this algebra. 
If one carries through the whole 
construction once more starting with $\omega_1$ 
instead of $\hat \phi$, then one gets a product state 
$\hat {\omega}_p$
for ${\cal R}_\beta (\O_\circ)$ and ${\cal R}_\beta (\O_3)'$ 
which coincides with the vector state induced by $\Omega_\beta$ on each algebra separately.  
\vskip .2cm
By a suitable smoothing procedure in the time variable
we can now construct a faithful normal product state $\omega_h$ for ${\cal R}_\beta (\O) 
\vee {\cal R}_\beta ( \hat {\cal O})'$ such that $\omega_h$ coincides with   
the vector state induced by $\Omega_\beta$ on both algebras:
Let $\hat{\chi}$ denote the
vector in the natural positive cone 
\# { {\cal P}^\natural_{\Omega_\beta } \bigl({\cal R}_\beta (\O_\circ) 
\vee {\cal R}_\beta (\O_3)'\bigr)  }
which induces $\hat {\omega}_p$ on ${\cal R}_\beta (\O_\circ) 
\vee {\cal R}_\beta (\O_3)'$ (see once again [BR, 2.5.31]). 
% Since $\hat{\chi) $ is 
% cyclic and separating for $\bigl({\cal R}_\beta (\O_\circ) 
% \vee {\cal R}_\beta (\O_3)'\bigr)$, $\hat{\chi) $ is not orthogonal to 
% $\Omega_\beta$.
It follows that there exists an isometry $I$ which satisfies
\# { I T \Omega_\beta =  T \hat{\chi}, 
\qquad \forall T \in {\cal R}_\beta (\O_3)' .}
Thus $I \in {\cal R}_\beta (\O_3)$ and 
$I \Omega_\beta \in {\cal D} 
({\rm e}^{- \lambda H_\beta})$ for all $0 \le \lambda \le \beta /2$.
This property implies that for any non-zero operator 
$C \in {\cal R}_\beta (\O_\circ) 
\vee {\cal R}_\beta (\O_3)'$
the set 
\# { \{ t \in \R : C {\rm e}^{it H_\beta} I \Omega_\beta \ne 0 \} }
is dense in $\R$. The details are as follows: assume there exists  
some interval $]t_1, t_2[$ such that
\# { C {\rm e}^{it H_\beta} I \Omega_\beta = 0  \qquad \forall t \in ]t_1, t_2[.}
The vector-valued function
\# { z \mapsto C {\rm e}^{iz H_\beta} I \Omega_\beta, \qquad 0 < \Im z < \beta /2,}
is analytic in the strip $0 < \Im z < \beta /2$ and continuous for $\Im z \searrow 0$.
Thus (141) implies that the function defined in $(142)$ vanishes identically. 
By assumption, $\Omega_\beta$ is the unique --- up to a phase --- time invariant
vector in $\H_\beta$.  
Taking an appropriate mean over the real axis we find
\# { 0  =    C \Omega_\beta \cdot ( \Omega_\beta \, , \, I \Omega_\beta ).}
Now $I \in {\cal R}_\beta (\O_3)$,  
$\omega_\beta$ is faithful for ${\cal R}_\beta (\O_3)$ and 
$\Omega_\beta$ is separating for
${\cal R}_\beta (\O_\circ) \vee {\cal R}_\beta (\O_3)'$. Therefore (143) implies $C = 0$ 
in contradiction to the assumption that $C$ is non-zero. Therefore the set 
(140) is dense in $\R$. 
Now let 
$h \in L_1 (\R)$ be a smooth positive function with 
support $] - \delta / 3 , \delta / 3 [$ and $\| h \|_1 = 1$. 
Locality together with (126) implies that
\# { {\cal R}_\beta (\O) \vee {\cal R}_\beta ( \hat {\cal O})' \ni C 
\mapsto \omega_h (C) =  \int_{- \delta / 3}^{\delta / 3} {\rm d}t \, h(t)  
(\hat{\chi} \, , \, {\rm e}^{-i t H_\beta} C {\rm e}^{i t H_\beta} \hat{\chi} ) }
defines a  product state for the pair ${\cal R}_\beta (\O)$ and
${\cal R}_\beta (\hat{\O})'$. In  fact,
\& {\omega_h  (AB) & =  
  \int_{- \delta / 3}^{\delta / 3} {\rm d}t \, h(t)  
(\Omega_\beta \, , \, {\rm e}^{-i t H_\beta} A {\rm e}^{i t H_\beta} \Omega_\beta)
(\Omega_\beta \, , \, {\rm e}^{-i t H_\beta} B {\rm e}^{i t H_\beta} \Omega_\beta) 
\cr
& = (\Omega_\beta \, , \, A \Omega_\beta)
(\Omega_\beta \, , \, B \Omega_\beta) ,
\quad \hbox{for} \quad A \in {\cal R}_\beta (\O) , 
\quad B \in {\cal R}_\beta ( \hat {\cal O})'. }
Thus the restriction $\omega_p$ of $\omega_h$ to the algebra 
${\cal R}_\beta (\O) \vee {\cal R}_\beta ( \hat {\cal O})'$ is independent of $h$ and
coincides with the vector state induced by $\Omega_\beta$ on both algebras.
Moreover, combining (140) and (144) we conclude that $\omega_p$ is faithful on
${\cal R}_\beta (\O) \vee {\cal R}_\beta ( \hat {\cal O})'$. 
\vskip .2cm
\noindent
(iv) $\Rightarrow$ (v) From [BR, 2.5.31] we infer that
there exists a unique vector $\eta $ in the natural positive 
cone ${\cal P}^\natural_{\Omega_\beta} \bigl({\cal R}_\beta (\O) \vee 
{\cal R}_\beta ( \hat {\cal O})' \bigr) $
such that 
\# { (\eta \, , \, C \eta) =   \omega_p (C) 
\qquad \forall C \in 
{\cal R}_\beta (\O) \vee {\cal R}_\beta ( \hat {\O})' .}
Moreover, $\omega_p$ is  faithful on
${\cal R}_\beta (\O) \vee {\cal R}_\beta ( \hat {\cal O})'$. Thus
$\eta$ is cyclic and separating 
for ${\cal R}_\beta (\O) \vee {\cal R}_\beta ( \hat {\cal O})'$.
\vskip .2cm
\noindent
(v) $\Rightarrow$ (vi)
Let $W_\eta$ be given by linear extension of 
\# { W_\eta A B \eta = A \Omega_\beta \otimes B \Omega_\beta. } 
Because of  (v) (b) $W_\eta$ is densely defined and isometric. Due to the Reeh--Schlieder
property of the KMS vector $\Omega_\beta$ the range of 
$W_\eta$ is dense in $\H_\beta \otimes \H_\beta$ too. 
Thus $W_\eta$ can be extended to a unitary operator 
$W 
\colon \H_\beta \to \H_\beta \otimes \H_\beta$. From (147) we infer
\# { W A B W^* = A  \otimes B   } 
for all $ A \in {\cal R}_\beta (\O) $ and $B \in {\cal R}_\beta ( \hat {\cal O})'$.
\vskip.2cm
\noindent
vi) $\Rightarrow$ i) This part has been provided in the proof of Theorem 4.11.}

\Rem{Property (vi) implies that the state $\omega_p$ specified in 
(iv) is uniquely fixed by the factorization property
\# {\omega_p  (AB)  =  
(\Omega_\beta \, , \, A \Omega_\beta)
(\Omega_\beta \, , \, B \Omega_\beta) ,
\qquad \forall A \in {\cal R}_\beta (\O) , \quad B \in {\cal R}_\beta ( \hat {\cal O})'. }
}

The split property has many interesting implications
which will be discussed in our next paper; here we will only quote one more
result of Buchholz [Bu]:

\Cor{Assume the inclusion 
${\cal R}_\beta (\O) \subset {\cal R}_\beta (\hat{\O})$
is split and
let $\O_a$ and $\O_b$ denote two regions contained in $\O$. If $\Phi$ is an isomorphism which
maps ${\cal R}_\beta (\O_a)$ onto ${\cal R}_\beta (\O_b)$, 
then $\Phi$ can be implemented by a unitary
operator $U \in {\cal N_\beta}$:
\# { \Phi (A) = U A U^{-1} .}
Hence $\Phi$ acts trivially on ${\cal R}_\beta (\hat{\O})'$.} 

\Pr{Once the existence of a cyclic and separating product vector
has been shown for ${\cal R}_\beta (\O)$
and ${\cal R}_\beta (\hat{\O})'$, Buchholz's result follows by the original arguments. 
We present them here for completeness only. 
Let $\eta$ denote the product vector specified in Theorem 5.1. v.) and
$P_1$ the projection onto $\overline { {\cal R}_\beta (\O) \eta } \subset \H_\beta$.
Clearly, $\overline { {\cal R}_\beta (\O) \eta }$ is invariant under the action of 
${\cal N}_\beta$. Thus we can consider the induced representation $\pi_{P_1}$ of
${\cal N}_\beta$ on $\overline { {\cal R}_\beta (\O) \eta }$. Since $\eta \in P_1 \H_\beta$,
this representation is faithful and it is easy to verify that
\# { \pi_{P_1} ({\cal N}_\beta) = \B (P_1 \H_\beta) . }
Now $\pi_{P_1} \bigl({\cal R}_\beta (\O_a)\bigr) \subset \pi_{P_1} ({\cal N}_\beta)$ and
$\pi_{P_1} \bigl({\cal R}_\beta (\O_b)\bigr) \subset \pi_{P_1} ({\cal N}_\beta)$ both have 
a cyclic vector,
namely $P_1 \Omega_\beta \in P_1 \H_\beta$ and a separating vector, namely
$\eta \in P_1 \H_\beta$.
Hence every isomorphism which maps $\pi_{P_1} \bigl({\cal R}_\beta (\O_a)\bigr)$ onto
$\pi_{P_1} \bigl({\cal R}_\beta (\O_b)\bigr)$ is spatial [Di p.222, Theorem 3]. 
Thus there exists a unitary operator $U \in \pi_{P_1} ({\cal N}_\beta)$ such that
\# { \pi_{P_1} \circ \Phi (A) = U \pi_{P_1}(A) U^{-1} 
\qquad \forall A \in {\cal R}_\beta (\O_a),}
and from this relation the statement follows immediately.}

\vskip .5cm
\Hl{Some more Remarks}

\noindent
Let $\eta$ denote the product vector constructed in Theorem 5.1.\ v.).
The set 
\# {  {\cal L}_\beta (\O, \hat{\O}) := \overline{ {\cal R}_\beta (\O) \eta  } }
is a convenient linear subset of the set of strictly localized thermal excitations 
${\cal L}_\beta (\hat{\O})$ defined in (25). In fact (see [BJ b]), 
\vskip .3cm
\halign{ \indent #  \hfil & \vtop { \parindent = 0pt \hsize=34em
                            \strut # \strut} 
\cr 
(i)    & ${\cal L}_\beta (\O, \hat{\O})$ is 
a closed subspace of $\H_\beta$;
\cr
(ii)    & ${\cal L}_\beta (\O, \hat{\O})$ is invariant under the action of 
${\cal R}_\beta (\O)$;
\cr
(iii)    & The vectors of ${\cal L}_\beta (\O, \hat{\O})$ induce product states on 
${\cal R}_\beta (\O) \vee {\cal R}_\beta (\hat{\O})'$, which coincide with the vector state 
induced by the KMS vector $\Omega_\beta$ on ${\cal R}_\beta (\hat{\O})'$: 
if $\Psi \in {\cal L}_\beta (\O, \hat{\O})$, then  
\# {(\Psi \, , \, AB \Psi) = (\Psi\, , \, A \Psi)  ( \Omega_\beta \, , \, B \Omega_\beta) }
for all  
$A \in {\cal R}_\beta (\O)$ and $B \in {\cal R}_\beta (\hat{\O})'$.
\cr
(iv)    & ${\cal L}_\beta (\O, \hat{\O})$ is complete in the following sense: 
to every normal state $\phi$
on ${\cal R}_\beta (\O)$ there exists a $\Phi \in {\cal L}_\beta (\O, \hat{\O})$ such that
\# {(\Phi\, , \, A \Phi) = \phi(A)  }
for all $A \in {\cal R}_\beta (\O)$. 
\cr
}  
\vskip .3cm
\noindent
Property (iv) can be seen as follows:
Since ${\cal R}_\beta (\O)$ has a cyclic and separating vector, there exists a vector 
$ \tilde {\Phi} \in \H_\beta$ which induces the given normal state
$\phi$ on  ${\cal R}_\beta (\O) $. Using the isomorphism specified in (119) we find that
$\Phi := W^* ( \tilde {\Phi} 
\otimes  \Omega_p )  \in \H_\Lambda $ satisfies (155).
\vskip 1cm

It was noticed by Buchholz, D'Antoni and Longo
that the split property imposes certain restrictions on the energy level density
of excitations of the KMS state described by the state vectors of
${\cal S}_\beta (\O, \lambda)$ [BD'AL b]:

\Th{Consider a TFT, specified by a
von Neumann algebra ${\cal R}_\beta$ with a cyclic and separating
vector $\Omega_\beta$ and a net of subalgebras
$\O \to {\cal R}_\beta (\O)$,  
subject to the conditions (i) and (ii) stated on p.\ 6.
Assume the inclusion 
${\cal R}_\beta (\O) \subset {\cal R}_\beta (\hat{\O})$
is split. Then the maps
\# {\matrix {\Theta_{\lambda, {\cal O} } \colon & {\cal R}_\beta (\O) & \to & \H_\beta \cr
& A & \mapsto & {\rm e}^{- \lambda H_\beta} A   \Omega_\beta , }} 
are compact for $0 < \lambda < \beta / 2 $. I.e., the set  
\# { {\cal S}_\beta (\O, \lambda) := \{ {\rm e}^{- \lambda | H_\beta |} A \Omega_\beta :
A \in {\cal R}_\beta (\O) , \| A \| \le 1 \} }
is relatively compact in the norm topology for all $\lambda > 0$.}

\Pr{The first statement is a consequence of [BD'AL b, Proposition 4.2.]
and [BD'AL b, Lemma 3.1.i).]. 
The second statement follows from an arguments,
 which we have already reproduced in the proof of Lemma 4.1.,
and which is also due to Buchholz, D'Antoni and Longo.} 
 
\Rem{As pointed out in [BD'AL b], it is clear that these limitations cannot be relaxed. Since
\# { {\rm e}^{-  {\beta \over 2}  H_\beta } A \Omega_\beta
= J A^*\Omega_\beta,}
the map $\Theta_{\lambda, {\cal O}}$ is not even compact for $\lambda = \beta /2$.}

Our nuclearity condition relies on decent infrared properties of the
generator of the time evolution. Our arguments are less conclusive, if $\omega_\beta$
describes a physical system at a critical point. But if 
the split property holds in the vacuum sector, then 
it holds also in the GNS representation associated with any thermal state 
which is locally normal w.r.t.\ the vacuum representation. Thus even
at a critical point the maps $\Theta_{\lambda, {\cal O} }$ should at least be
compact for $0 < \lambda < \beta / 2 $, as long as the corresponding 
KMS state is locally normal w.r.t.\ the vacuum representation. However, there is 
the possibility that infrared divergencies might destroy local normality 
(see e.g.\ [BJ a][BR, Ex.\ 5.4.15]). Despite the
general belief that in 3+1 space--time dimensinons
all states of physical interest should be locally normal 
to each other, we can not rule out this possibility.

% The split property implies that $\H_\beta$ is separable.
% If $\H_\beta$ fails to be separable, 
% as may be expected for a general KMS state at a critical point,
% then the split property can not hold.
% If the split property nevertheless holds in the vacuum representation then 
% $\Omega_\beta$ can not be locally normal w.r.t.\ the vacuum representation.
% This is in agreement with our physical picture: close to a critical point there will be many
% diverse states which can be prepared by local operations with small energy--momentum
% transfer. One might think of a bit of dust thrown in a 
% coexistence phase point of a gaseous and a liquid phase.
% Thus if the split property fails in the GNS representation associated with $\omega_\beta$
% and at the same time holds in the vacuum representation, then $\omega_\beta$ can not be
% locally normal w.r.t.\ the vacuum representation.  
% Although there is a general belief that all states of physical interest should be locally normal 
% to each other, the possibility that infrared divergencies might destroy local normality 
% is  well known (see e.g.\ [BJ a][BR, Ex.\ 5.4.15]). Thus we expect 

\vskip 1cm
\noindent
{\it  Acknowledgements.\/}
\noindent
The author is indebted to D.\ Buchholz for helpful discussions concerning the nuclearity condition 
as well as the proof of the split property. 
W.~Thirring has always stressed the importance of sharp bounds in his lectures.
The present work demonstrates how the algebraic structure of TFT rests on bounds for 
the correlation functions. This work was supported by the CEE and 
the Fond zur F\"orderung der Wissenschaftlichen Forschung in \"Osterreich, 
Proj.\ Nr.\ P10629 PHY.  

\vskip 1cm

\noindent
{\fourteenrm References}
\nobreak
\vskip .3cm
\nobreak
\halign{   &  \vtop { \parindent=0pt \hsize=33em
                            \strut  # \strut} \cr 
\REF
{Bo}
{Borchers, H.J.}     {A remark on a theorem of B.\ Misra}
                    	{\CMP} 
                     {4}   {315--223}
                     {1967}
\REF
{Bu}
{Buchholz, D.}       {Product states for local algebras}
                    	{\CMP} 
                     {36}   {287--304}
                     {1974}
\REF
{BB}
{Bros, J.\ and Buchholz, D.}      {Towards a relativistic KMS-condition}
                                  {Nucl.\ Phys.\ B} 
                                  {429} {291--318}
                                  {1994}
% \REF
% {BD'AF}
% {Buchholz, D., D'Antoni, C.\ and Fredenhagen, K.}
%                                    {The universal structure of local algebras}
%                                    {\CMP}
%                                    {111}  {123--135} 
%                                    {1987}
\REF
{BD'AL a}
{Buchholz, D., D'Antoni, C.\ and Longo, R.}
                                    {Nuclear maps and modular structure I:
                                     General properties}
                                    {J.\ Funct. Analysis}
                                    {88/2}  {233--250} 
                                    {1990}
\REF
{BD'AL b}
{Buchholz, D., D'Antoni, C.\ and Longo, R.}
                                    {Nuclear maps and modular structure II:
                                     Applications to quantum field theory}
                                    {\CMP}
                                    {129}  {115--138} 
                                    {1990}
\REF
{BDL}
{Buchholz, D., Doplicher, S.\ and Longo, R.}
                                    {On Noether's theorem in quantum field theory}
                                    {Ann.\ Phys.}
                                    {170}  {1--17} 
                                    {1986}
\BOOK
{BR}  
{Bratteli, O.\ and Robinson, D.W.} {Operator Algebras and Quantum Statistical Mechanics~I,II} 
                                  {Sprin\-ger-Verlag, New York-Heidelberg-Berlin} 
                                  {1981}
\REF
{BJ a}
{Buchholz, D.\ and Junglas, P.}   {Local properties of equilibrium states and the particle
                                   spectrum in quantum field theory} 
                                  {Lett.\ Math.\ Phys} 
                                  {11} {51--58}
                                  {1986}
\REF
{BJ b}
{Buchholz, D.\ and Junglas, P.}   {On the existence of equilibrium states in local 
                                   quantum field theory} 
                                  {\CMP} 
                                  {121} {255--270}
                                  {1989}
\REF
{BV}
{Buchholz, D.\ and Verch, R.}   {Scaling algebras and renormalization group in 
                                 algebraic quantum field theory} 
                                  {\RMP} 
                                  {7/8} {1195--1239}
                                  {1995}
\REF
{BW}
{Buchholz, D.\ and Wichmann, E.}   {Causal independence and the energy-level
                                   density of states in local quantum field theory} 
                                  {\CMP} 
                                  {106} {321--344}
                                  {1986}
\REF
{BY}
{Buchholz, D., Yngvason, J.}       {Generalized nuclearity conditions and the split
                                    property in quantum field theory}
                                             	{Lett.\ Math.\ Phys} 
                                              {23}   {159--167}
                                              {1991}
% \REF
% {D'A}
% {D'Antoni, C.}
%                       {Technical properties of the quasi-local algebra}
%                       {The algebraic theory of superselection sectors. Introduction and recent
%                        results.  }{Proceedings of the convegno internationale algebraic 
%                        theory of superselection sectors and field theory, Palermo
%                        Nov.\ 23--30, 1989} 
%                       {Ed.: D.\ Kastler} {World Scientific}
%                       {1990}  
% {\sevenbf [D'ADFL]}  & \hskip -9.5cm \vtop {
%                    {\sevenrm D'Antoni, C., Doplicher, S., Fredenhagen, K., and Longo, R.,} 
%                    {\sevensl Convergence of local charges and continuity properties of 
%                        $\scriptstyle W^*$-inclusions,} 
%                                {\sevenrm \CMP} 
%                                {\sevenbf 110,} 
%                                {\sevenrm 321--344} 
%                                {\sevenrm (1987)}
%                                {\sevensl and} 
%                                {\sevenrm Erratum} 
%                                {\sevenrm \CMP} 
%                                {\sevenbf 116,} 
%                                {\sevenrm 321--344}
%                                {\sevenrm (1988)} } \cr
% \REF
% {D'AL}
% {D'Antoni, C.\ and Longo, R.}       {Interpolation by type I factors and the flip automorphism}
%                                   {J.\ Funct.\ Analysis}
%                                   {51}    {361--371}
%                                   {1983}
\REF
{DL}
{Doplicher, S.\ and Longo, R.}      {Standard and split inclusions of von Neumann algebras}
                                   {Invent.\ Math.}
                                   {73}    {493--536}
                                   {1984}
\BOOK
{Di}
{Dixmier, J.}    {$\scriptstyle C^*$-Algebras and their Representations} 
              {North Holland} 
              {1990}
\REF
{FS}
{Florig, M.\ and Summers, S.J.}      {On the statistical independence of algebras of observables}
                       {\JMP}
                       {38/3} {1318--1328}
                       {1997}
\BOOK
{H}
{Haag, R.}    {Local Quantum Physics: Fields, Particles, Algebras} 
              {Springer-Verlag, Berlin-Heidelberg-New York} 
              {1992}
\REF
{HHW}
{Haag, R., Hugenholtz, N.M.\ and Winnink, M.}
                          {On the equilibrium states in quantum statistical mechanics}  
                          {\CMP}	
                          {5}	{215--236}
                          {1967}
\REF
{HS}
{Haag, R.\ and Swieca, J.A.}  {When does a quantum field theory describe particles ?}
                             {\CMP}	
                             {1} {308--320}
                             {1965}
\BOOK
{Ja}
{Jarchow, H.}     {Locally Convex Spaces} 
                  {Stuttgart, Teubner}  
                  {1981}
\HEP
{J\"a a}
{J\"akel, C.D.}                    {Two algebraic properties of thermal quantum field theories}
						                             {to be published in J.\ Math.\ Phys}
\HEP
{J\"a b}
{J\"akel, C.D.}                    {The Reeh--Schlieder property for thermal field theories}
						                             {to be published in J.\ Math.\ Phys}
\REF
{J\"a c}
{J\"akel, C.D.}                   {Decay of spatial correlations in thermal states}
                                  {Ann.\ l'Inst.\ H.\ Poincar\'e} 
                                  {69}	{425--440}
                                  {1998}
% \HEP
% {J\"a c}
% {J\"akel, C.D.}                   {On the relation between KMS states for different temperatures}
%                                  {hep-th/9803245} 
% \HEP
% {J\"a d}
% {J\"akel, C.D.}                    {Cluster estimates for modular structures}
%                                    {hep-th/9804017} 
\BOOK
{Ju}
{Junglas, P.}   {Thermodynamisches Gleichgewicht und Energiespektrum in der 
                 Quantenfeldtheorie} 
                {Dissertation, Hamburg}
                {1987}
\BOOK
{P}
{Pietsch, A.}     {Nuclear Locally Convex Spaces} 
                  {Springer-Verlag, Berlin-Heidelberg-New York}  
                  {1972}
\BOOK
{Sa}
{Sakai, S.}     {$\scriptstyle C^*$-Algebras and $\scriptstyle W^*$-Algebras} 
                     {Springer-Verlag, Berlin-Heidel\-berg-New York}  
                     {1971}
% \REF
% {Sch}
% {Schlieder, S.}                     {Einige Bemerkungen \"uber Projektionsoperatoren}
%                                    {\CMP}
%                                    {13} {216--225}
%                                    {1969}
\REF
{Su}
{Summers, S.J.}       {On the statistical independence of algebras of observables}
                       {\RMP}
                       {2/2} {201--247}
                       {1990}
\BOOK
{Ta}
{Takesaki, M.}   {Theory of Operator Algebras I}
                 {Springer-Verlag, Berlin-Heidel\-berg-New York}  
                 {1979}
\cr}

\bye